\documentclass[12pt,preprint]{aastex}

\newcommand\msol{{\cal M_{\odot}}}

\newcommand\mstar{{\cal M}_*}

\newcommand\teff{{T_{\rm eff}}}
\newcommand\amlt{{\alpha_{\rm MLT}}}

\newcommand\fttau{T(\tau,\,T_{\rm eff})}
\newcommand\lta{\mathrel{\hbox{\raise 0.6 ex \hbox{$<$}\kern
                   -1.8 ex\lower .5 ex\hbox{$\sim$}}}}
\newcommand\gta{\mathrel{\hbox{\raise 0.6 ex \hbox{$>$}\kern
                   -1.7 ex\lower .5 ex\hbox{$\sim$}}}}
 
\shortauthors{VandenBerg et al.}
\shorttitle{Stellar Models With Blanketed Atmospheres}
 
\begin{document}
 
\title{On the Use of Blanketed Atmospheres as Boundary Conditions for
Stellar Evolutionary Models}
 
\author{Don A.~VandenBerg}
\affil{Department of Physics \& Astronomy, University of Victoria,
       P.O.~Box 3055, Victoria, B.C., V8W~3P6, Canada}
\email{vandenbe@uvic.ca}

\author{Bengt Edvardsson}
\affil{Uppsala Astronomical Observatory, Box 515, SE-751 20 Uppsala, Swenden}
\email{be@astro.uu.se}

\author{Kjell Eriksson}
\affil{Uppsala Astronomical Observatory, Box 515, SE-751 20 Uppsala, Sweden}
\email{Kjell.Eriksson@astro.uu.se}

\author{Bengt Gustafsson}
\affil{Uppsala Astronomical Observatory, Box 515, SE-751 20 Uppsala, Sweden}
\email{bg@astro.uu.se}
 
\begin{abstract}
Stellar models have been computed for stars having [Fe/H] $= 0.0$ and $-2.0$ to
determine the effects, primarily on the predicted $\teff$ scale, of using
boundary conditions derived from the latest MARCS model atmospheres.  The latter
have been fitted to the interior models at both the photosphere and at $\tau =
100$, and, at least for the $0.8$--$1.0 \msol$ stars considered here, the
resultant evolutionary tracks on the H-R diagram were found to be nearly
independent of the chosen fitting point.  Particular care was taken to treat
the entire star as consistently as possible; i.e., both the stellar structure
and atmosphere codes assumed the same abundances of helium and the heavy
elements, as well as the same treatment of convection (the mixing-length theory,
with the free parameter $\amlt$ chosen to satisfy the solar constraint).
Tracks were also computed in which the photospheric pressure was obtained by
integrating the hydrostatic equation together with either the classical gray
$\fttau$ relation or that derived by Krishna Swamy (1966) from an
empirical solar atmosphere.  Due to the compensating effects of differences in
the calibrated values of $\amlt$, the evolutionary sequences that assumed gray
atmospheric structures were in fortuitously close agreement with those using
MARCS atmospheres as boundary conditions, independently of the assumed
metallicity.  (The structures of gray atmospheres are quite different from
those predicted by MARCS models.)  On the other hand, the models based on the
Krishna Swamy $\fttau$ relationship implied a much hotter giant branch (by
$\sim 150$ K) at solar abundances, which happens to be in good agreement with
the inferred temperatures of giants in the open cluster M$\,$67 from the latest
$(V-K)$--$\teff$ relations, though they were similar to the other cases at
[Fe/H] $= -2.0$.  Most of the computations assumed $Z = 0.0125$ for the Sun, as
derived by M.~Asplund and colleagues, though a few models were calculated for
$Z = 0.0165$, assuming the Grevesse \& Sauval (1998) metals mix, to determine
the dependence of the evolutionary tracks on $Z_\odot$.  Grids of ``scaled
solar, differentially corrected" (SDC) atmospheres were also computed, to try to
improve upon theoretical MARCS models.  When they were used to describe the
surface layers of stellar models, the resultant tracks were in remarkably good
agreement with those that employed a standard scaled-solar (e.g., Krishna Swamy)
$\fttau$ relation to derive the boundary pressures, independently of the
assumed metal abundance.  To within current distance and metallicity
uncertainties, it was possible to obtain a good match of isochrones for [Fe/H]
$=-2.0$ to the C-M diagram of the globular cluster M$\,$68.  Until these
uncertainties and those associated with the atmospheric boundary conditions are
reduced significantly, it cannot be claimed with any confidence that $\amlt$
does or does not vary with [Fe/H].  While our consideration of M$\,$67 giants
suggests that this parameter is independent of $\teff$ and $\log g$, some
(small) variation with stellar parameters cannot be ruled out.
\end{abstract}
 
 
\section{Introduction}
\label{sec:intro}
Nearly all published stellar models, including the large grids of evolutionary
calculations by \citet{gbb00}, \citet{ydk01}, \citet{pcs04}, and \citet{vbd06},
have described the atmospheric layers using either the $\fttau$ relation
from an empirical solar atmosphere ({\citealt{ks66}; hereafter KS66 --- see his
equation 33), or that obtained for a gray atmosphere when the Eddington
approximation is assumed.\footnote{In this investigation, the Greek letter
$\tau$ is used, without exception, to represent the Rosseland mean optical
depth.  In addition, the term ``photosphere" is used here to refer to that layer
in a star where the local temperature equals the effective temperature.  (This
definition is generally employed by those who construct stellar interior
models.)  Thus, it should be understood that, e.g., the ``photospheric pressure"
is the pressure at $T=\teff$.}  As noted by, e.g., \citet{van91} and
\citet{scw02}, just this difference alone can affect the predicted $\teff$ scale
by up to $\sim 100$ K.  Consequently, models that treat the atmosphere
differently must also adopt different values for the mixing-length parameter,
$\amlt$, in order to satisfy the solar constraint.  This, in turn, will have
important ramifications for the predicted temperatures of giants, in particular,
since the location of the red giant branch (RGB) on the H-R diagram is known to
be a sensitive function of $\amlt$ (see, e.g., \citealt{van83}).  In view of the
critical role played by the atmosphere in calculations of stellar structure and
evolution, it is obviously important to treat the outermost layers as accurately
as possible.

The most straightforward way of avoiding, or at least minimizing, the systematic
errors that must occur when a fixed $\fttau$ relation is used to describe the
temperature distribution in the surface layers is instead to attach modern model
atmospheres onto interior structures.  In this way, the differential effects of
changes in the fundamental atmospheric parameters ($\teff$, surface gravity, and
chemical composition) are hopefully incorporated in the predicted $T$--$\tau$
structures.  Preferably, atmospheres should be attached to interior models at
large optical depths where the diffusion equation, which is the limiting form
of the transfer equation that is solved by a stellar structure code, is valid.
By comparing the radiative flux given by the diffusion equation with the exact
values from model atmospheres, \citet{mvp94} showed that the two determinations
agree only at $\tau\gta 10$ in the case of solar-type stars, and at $\tau\gta
20$ in stars of very low gravity.  However, the impact of matching atmosphere
and interior models at smaller optical depths appears to be quite minor.
\citet{mkd01} reported that evolutionary tracks for $0.8 \msol$ stars having
$Z=0.0002$ differed only slightly when model atmospheres were joined to interior
models at $\tau =1$, 10, or 100.  Much bigger effects on predicted temperatures
were found if convection was treated differently in the atmosphere than in the
interior, which suggests that it is very important to model the entire star as
consistently as possible.

In fact, most studies to date of the consequences of using model atmospheres as
boundary conditions have not had complete consistency between the atmosphere
and interior models, insofar as the adopted physics and/or chemical abundances
are concerned.  This includes the work by \citet{vsr00}, who found that
evolutionary tracks for metal-deficient $0.8 \msol$ stars were offset to warmer
temperatures, by amounts that increased with decreasing [Fe/H], when MARCS
model atmospheres from the late 1990s were used to describe the surface layers
instead of the KS66 $\fttau$ relationship.  \citet{scw02} showed that giant
branches utilizing Kurucz model atmospheres tended to be steeper than those
obtained on the basis of gray or KS66 $T$--$\tau$ structures, but it was also
pointed out that the treatment of convection in the atmospheres that they
employed was not the same as in their interior models.  The same inconsistency
was acknowledged to be a concern in the study of $\alpha$-element enhanced
stellar models by \citet{csc04}.  Whether or not the noted inconsistencies
have a large or small effect on the comparisons that were presented in those
investigations is not known.

A follow-up study is clearly warranted --- one in which the atmosphere and
interior models assume exactly the same treatment of convection, the same
abundances of helium and the most important heavy elements, and (as much as is
practically possible) the same opacities and thermodynamics.  In order to
explore the consequences of fully consistent atmosphere-interior models for, in
particular, the predicted temperatures of stars, the Victoria stellar structure
code (see VandenBerg et al.~2006, and references therein) has been modified
to use model atmospheres produced by the latest version of the MARCS code (e.g.,
see \citealt{gee03}\footnote{Many improvements have been to this computer
program since 2003: they will be described in a forthcoming paper.  Note, as
well, that MARCS model atmospheres cooler than 4000 K are still in development
at the present time; consequently, evolutionary models for stars with masses
$\lta 0.6 \msol$ are not considered in this investigation.  However, reference
may be made to, e.g., \citet{bca97} and \citet{bcc98} for some discussion of
models for very low mass stars that take recent model atmospheres into
account.}) as boundary conditions.  As shown in \S~\ref{sec:methods}, which is
primarily concerned with numerical methods, the physical quantities predicted
by these two codes are in excellent agreement when the same chemical abundances
are assumed. \S~3 presents
several evolutionary sequences applicable to the Sun wherein the
surface layers are treated in three different ways; namely, MARCS atmospheres,
as well as the often-used gray and scaled-solar $\fttau$ descriptions (e.g.,
KS66).  This section also examines the impact on solar tracks of adopting
either the new solar abundances derived by M.~Asplund and collaborators from
3-D, non-LTE analyses of the solar spectrum (for a summary, see \citealt{ags06})
or the standard abundance distribution that was published by \citet{gs98} prior
to the aforementioned work.  In addition, the inferred temperatures of M$\,$67
giants from $V-K$ photometry suggest a preference for MARCS model atmospheres
that have been corrected semiempirically so as to better represent the solar
atmosphere.  \S~4 contains a similar analysis of the effects of variations in
the treatment of the atmosphere for the tracks of metal-poor stars, and for
isochrones appropriate to the globular cluster M$\,$68, while concluding
remarks are given in \S~5.
 
\section{Numerical Methods and Consistency Tests}
\label{sec:methods}

\subsection{Convection, Solar Metal Abundances, Opacities}
\label{subsec:physics}

Although the local mixing-length theory (MLT) of convection ({\citealt{bv58}) is
employed by both the MARCS model atmosphere and the Victoria stellar evolution
codes, a consistent treatment of convection would {\it not} be obtained if both 
computer programs adopted the same value of the free parameter, $\amlt$.  This
is for the reason that the most general form of the MLT (\citealt{hvb65})
contains many other parameters, including (in particular) $y$ and $\nu$, which
describe the temperature distribution and the energy dissipation of a convective
bubble, respectively.  (Henyey et al.~also define a quantity $\beta$, which is
used to determine the turbulent pressure from the convective velocity, and they
suggest using a correction to the diffusion equation denoted by $f$.)  Whereas
the MARCS code has generally adopted $y=0.076$ and $\nu = 8$ over the years, the
stellar models generated by the Victoria code have nearly always assumed $y =
{1\over 3}$ and $\nu = 8$.  As shown by \citet{pvi90}, it is possible to
compensate for a difference in $y$ (or $\nu$, for that matter) by adjusting the
value of $\amlt$ appropriately.  However, the generalized version of the MLT is
included in the Victoria code, and hence $y$ can simply be set to the default
value used in the MARCS code (i.e., 0.076).  Both codes will then have the same
treatment of convection if they each assume identical values of $\amlt$.

The Sun is traditionally used to calibrate $\amlt$ because it is the only star
for which the fundamental parameters are sufficiently well determined.  However,
the value that is obtained for $\amlt$ will depend on the adopted metal
abundance, an initial helium content that is determined from the requirement
that a $1.0 \msol$ model for the assumed metallicity reproduces the solar
luminosity at the solar age, and the treatment of the atmosphere.  Stellar
evolution codes typically consider only the 19 metals listed in the first column
of Table~\ref{tab:tab1} because they, along with H and He, constitute the entire
set of elements for which OPAL Rosseland mean opacities (\citealt{ir96}) may
be computed for stellar interior conditions using the Livermore Laboratory
website facility.\footnote{See http://www-phys.llnl.gov/Research/OPAL/.} 

In a concurrent study, \citet{vge07} examined the fits of isochrones to the
M$\,$67 color-magnitude diagram, assuming the same chemical abundances adopted
in this study; i.e., the $\log N$ abundances (on the scale where
$\log N_{\rm H} = 12.0$) listed in the last two columns of Table~\ref{tab:tab1}.
These give, in turn, the solar abundances published by \citet{gs98} and the
distribution that is obtained when the revised abundances for several of the
elements, as derived by M.~Asplund and collaborators from analyses of the
photospheric spectrum using 3-D, non-LTE model atmospheres, are taken into
account.  (Note that the entries for C and N in the third column are higher by
0.02 dex than the estimates published by \citealt{ags06}).  The opacities
computed by VandenBerg et al.~are also used here --- not only OPAL data for
temperatures greater than $\log T \sim 3.8$, but also complementary opacities
for lower temperatures, which include the contributions from molecules and
grains.  The low-$T$ opacities were calculated for the same heavy-element
mixtures using the computer program described by \citet{faa05}.

\subsection{The Implementation of Atmospheric Boundary Conditions}
\label{subsec:atmbc}

Before discussing the calculation of Standard Solar Models, it is useful to
summarize how the Lagrangian version of the Victoria evolutionary code operates
(also see \citealt{van92}).  The usual stellar structure equations are solved
by the Henyey method (\citealt{hfg64}) only in the region of the star containing
the innermost 99\% of its mass.  Two boundary conditions: $$\log {\cal R} =
a_1\,\log P + a_2\,\log T + a_3$$ and $$\log {\cal L} = b_1\,\log P +
b_2\,\log T + b_3$$ (where ${\cal R}$ and ${\cal L}$, in units of $10^{11}$ cm
and $10^{33}$ erg/s, are the adopted radius and luminosity variables,
respectively) are added to the system of difference equations at ${\cal M} =
0.99\,\mstar$, while two additional boundary conditions are applied at, or near,
the star's center.  To determine the coefficients $a_i$ and $b_i$, Runge-Kutta
integrations of 3 of the 4 stellar structure equations (assuming that the
integrated luminosity is constant in the surface layers) are performed at three
points on the H-R diagram that enclose the approximate $L$ and $\teff$ of the
model under consideration.

To begin these integrations it is necessary to have initial values for the
temperature, radius, and pressure.  If the starting point is taken to be the
photosphere, then $T_{\rm initial} = \teff$, ${{\cal R}_{\rm initial}}$ can be
calculated from $L = 4{\pi}R^2{\sigma}\teff^4$, and $P_{\rm initial}$ may be
determined either by (i) integrating the hydrostatic equation (${dP\over d\tau}
= {g\over\kappa}$) from very small optical depths to the value of $\tau$ where 
$T=\teff$, on the assumption that the temperature distribution obeys the KS66
(or gray) $\fttau$ relation, or (ii) interpolating in tables of photospheric
pressures obtained from fully consistent model atmospheres that have been
computed for a sufficiently wide range in $\log g$ and $\teff$.  If model
atmospheres are attached at depth (e.g., at $\tau = 100$), then the
interpolations in the grids of model atmospheres must provide the initial values
of $T$, ${\cal R}$, and $P$ at the chosen fitting point.  (Because the flux has
a $1/{\rm r}^2$ dependence, the values of the physical variables at depth will
have some errors associated with them when the atmospheres are very thick, if
derived from 1-D, flux-constant model atmospheres.  However, as discussed in
\S~\ref{subsec:m67rgb}, this is not a concern for the stellar models presented
in this study.)

In practice, an additional equation must be defined at the end-point of the
Runge-Kutta integrations so that the radius of the stellar model may be
calculated once the Henyey scheme has converged: that equation is $\log r/R =
c_1\,\log P + c_2\,\log T + c_3$.  (As for the boundary conditions, three
integrations down to $0.99 {\cal M}_*$ yield the values of $\log r/R$,
$\log P$, and $\log T$ that are needed to define three equations involving the
unknown quantities, $c_1$, $c_2$, and $c_3$, which may be solved using standard
methods.)  As the model moves along its evolutionary track, the three points
on the H-R diagram are adjusted, as necessary (using, e.g., the triangle
strategy of \citealt{kwh67}), so that the triangle they define always encloses
the evolving model, and the ``surface" boundary conditions are recomputed.

\subsection{Standard Solar Models}
\label{subsec:solmod}

The calculation of Standard Solar Models that made use of MARCS model
atmospheres proceeded in the following way.  For initial estimates of the
values of $\amlt$ and $\log N_{\rm He}$ that apply to the Sun, a small grid of
model atmospheres (for effective temperatures ranging from 5500 K to 6250 K, in
steps of 250 K, and, at each $\teff$, $\log g$ values from 4.0 to 4.9, in steps
of 0.3 dex) was calculated.  An evolutionary track for a $1.0 \msol$ star was
then computed from the zero-age main sequence (ZAMS) to an age of 4.6 Gyr using
the MARCS atmospheres as boundary conditions (via the methods described above).
The solar-age model so obtained would generally fail to reproduce the properties
of the Sun (specifically, its luminosity and $\teff$) satisfactorily.  As a
result, the trial values of $\amlt$ and $\log N_{\rm He}$ were revised, the
small grid of model atmospheres was recomputed to be consistent with updated
values of these parameters, and a new solar track was calculated.  After a few
iterations, a solar model was obtained that reproduced the observed $\log L$
and $\log\teff$ values to within 0.0002 dex.

This procedure was repeated for each of the two heavy-element mixtures given
in Table~\ref{tab:tab1}, resulting in the values of $\amlt$, $\log N_{\rm He}$,
$X$, $Y$, and $Z$ contained in Table~\ref{tab:tab2}.  (The model atmospheres
and low-T opacities consider many more elements than those listed in
Table~\ref{tab:tab1}, but they do not affect the overall mass-fraction
abundances appreciably because their abundances are low.  Their effects on the
blanketing and opacity are, however, taken into account.) 

\subsection{Comparison of Results from the MARCS and Victoria Codes}
\label{subsec:compare}

To check the consistency of the MARCS and Victoria codes, both computer programs
were used to calculate the structure of the sub-photospheric layers of the Sun.
In Figure~\ref{fig:vanfig1}, the solid
curves indicate the predicted $T$, $P$, opacity, $\ldots$ profiles between the
photosphere and that layer where $\tau = 100$ as predicted by the MARCS
atmosphere code for a star having $\log g = 4.44$, $\teff = 5777$ K, the Asplund
mix of heavy elements, and the values of $\amlt$ and $\log N_{\rm He}$ given in
Table~\ref{tab:tab2}.  The dashed curves, on the other hand, represent the
results of Runge-Kutta integrations by the Victoria evolutionary code inward
from the solar photosphere for the same abundances and parameter values, and
assuming that the initial values of the dependent variables are the predicted
pressure at $T=\teff$ from the MARCS model, the observed $\teff$, and the
observed radius of the Sun.  The agreement between the two is clearly excellent,
which demonstrates that the effects of differences in the physics
implemented in the two codes are very minor.  (A similar comparison
was carried out for a model on the lower RGB having $\teff = 4726$ K and
$\log g = 3.45$, and the results were qualitatively almost identical.)

To be sure, some variations are expected because, in particular, the diffusion
equation approximation to the transfer equation is used in the Victoria code,
while the MARCS code solves the monochromatic form of the transfer equation for
$\sim 10^5$ wavelength points.  In fact, the small discrepancies in the
temperature profiles in Fig.~\ref{fig:vanfig1}, are consistent with this
difference.  Using the diffusion approximation together with Rosseland mean
opacities {\it should} lead to a shallower temperature gradient at $\tau\lta 1$
than that predicted by model atmospheres.  As discussed by, e.g., \citet{mi70},
a better approximation to the true gradient would be obtained if the larger
Planck means were used in the outer layers.  In any case, the level of
consistency between the MARCS and Victoria codes is clearly very satisfactory,
and we can be confident that the atmosphere and interior structures are about
as well matched as it is possible to make them.

\subsection{On the Mass Dependence of Model Atmospheres}
\label{subsec:massatm}

We conclude this section with a few comments on the question of whether a model
atmosphere has any direct dependence on the assumed mass in addition to its
dependence on surface gravity.  For a plane-parallel model, the mass is not
defined and the radius, in principle, is infinite.  Plane-parallel
stratification is a reasonable approximation when the atmospheric geometrical
depth is a very small fraction of the star's radius, as in dwarfs
and subgiants.  However, sphericity effects do introduce a mass dependence in
very extended models.  To illustrate this, spherical model atmospheres have
been computed for $\teff = 4000$ K, $\log g = 0.0$, $Z=0.05$, and masses of
0.5 and $5.0 \msol$ (which have total radii of 117 and $371 {{\cal R}_\odot}$,
respectively).  The predicted $T$--$\tau$, depth--$\tau$, and $T$--depth
structures for these two cases are shown in Figure~\ref{fig:vanfig2}.  Because
the atmosphere from $\tau \sim 10^{-4}$ to $\tau \sim 1$ is a much larger
fraction of the total radius in the $0.5 \msol$ star ($\approx 20$\%) than in
the $5.0 \msol$ star ($\approx 5$\%), the sphericity effects on the radiation
field are considerably larger in the lower mass star.  These cause reduced
temperatures in the outermost layers and a steeper temperature gradient over
most of the atmosphere, as evident in Fig.~2b. However, even in this fairly
extreme example, the structures of the inner atmospheres (from the photosphere
to $\tau = 100$, which is the main region of interest in this investigation)
do not differ by very much.  (For additional discussion concerning the effects
of varying atmospheric extension, see \citealt{he06}.)

\section{Models for [Fe/H] $= 0.0$ Stars}
\label{sec:solar}

Most of the following analysis of the impact of different treatments of the
atmospheric layers on the structure and evolution of solar abundance stellar
models assumes the Asplund mix of heavy elements (Table~\ref{tab:tab1}) and the
corresponding $X$, $Y$, and $Z$ values given in Table~\ref{tab:tab2}.  Models
for the Grevesse \& Sauval (1998; hereafter GS98) metals mixture are considered
(in a separate subsection) mainly to illustrate the dependence of evolutionary
tracks with consistent model atmospheres on the distribution of the elemental
abundances when the atmospheric layers are treated similarly (i.e., by fully
consistent MARCS models).  Note that all evolutionary calculations presented in
this paper assume that $\amlt$ is independent of mass, metallicity, and
evolutionary state, given that compelling evidence in support of any such
dependence has not yet been found (see the summary of work to date on this issue
by \citealt{fvs06}\footnote{These authors suggest that there {\it may} be a
difference in the value of $\amlt$ that must be assumed in stellar models to
reproduce the temperatures of globular cluster giants, as derived from $V-K$
photometry, and that required by a Standard Solar Model, {\it if} the new
Asplund metal abundances for the Sun are assumed.  Such inferences are highly
speculative because they are subject to the large uncertainties that persist in
the cluster [Fe/H] scale and the treatment of the atmospheric layers in stellar
models, among other things.  Ample support for this assertion is provided in
the present study.}).

\subsection{Solar Abundance Models, Assuming $Z_\odot=0.01247$ (Asplund)}
\label{subsec:Asplund}

Having determined the helium abundance and the value of $\amlt$ needed to fit
the Sun (see \S~\ref{subsec:solmod} and Table~\ref{tab:tab2}), it is a
straightforward exercise to produce evolutionary tracks for solar parameters
that extend to the lower RGB.  To achieve this, a much larger grid of model
atmospheres (for $4000\le\teff\le 8000$ K and $3.0\le\log g\le 5.0$, in steps
of 250 K and 0.5 dex, respectively) was computed, assuming the Asplund
abundances and the calibrated values of $\log N_{\rm He}$ and $\amlt$.  The
tracks for a $1.0 \msol$ model that were obtained when these atmospheres were
fitted to interior structures at the photosphere, on the one hand, and at
$\tau = 100$, on the other, are shown in Figure~\ref{fig:vanfig3} as solid and
dashed curves.  The two evolutionary sequences are nearly identical, which
implies that, at least for this case, the derived $\teff$ scale is essentially
independent of the fitting point.

Also plotted in Fig.~\ref{fig:vanfig3} are evolutionary tracks for the same
mass, $Y$, and $Z$ in which either the KS66 $\fttau$ relation (the dotted curve)
or that for a gray atmosphere (dot-dashed curve) is used to describe the
atmospheric layers exterior to the photosphere.  In both of these cases, the
mixing-length parameter was chosen so that the solar constraint was satisfied:
hence all of the tracks are coincident at the location of the solar symbol.
Interestingly, the track using gray atmospheres lies very close to those
employing MARCS model atmospheres, while the giant branch of the track using
KS66 atmospheres is hotter than the rest by $\sim 100$ K.  In fact, these
results are a consequence of the differences in the assumed values of $\amlt$.

This can be readily substantiated with the aid of Figure~\ref{fig:vanfig4},
which compares the evolutionary sequences that are obtained for the three
different treatments of the atmospheric layers considered in
Fig.~\ref{fig:vanfig3} when the same value of the mixing-length parameter is
assumed.  Whereas, to first order, changes in the treatment of the atmosphere
shift an entire track by a roughly constant amount in $\teff$, varying $\amlt$
has a 2--3 times larger impact on the temperatures of giants than on the
temperatures of main-sequence stars (see, e.g., Fig.~5 by \citealt{van91}).
In Fig.~\ref{fig:vanfig4}, the dot-dashed curve is closer to the solid curve
in the vicinity of the main sequence than along the RGB, while the opposite is
true in the case of the dotted curve.  It is, therefore, to be expected that
the dot-dashed curve will end up being just slightly cooler, and the dotted
curve considerably hotter, than the solid curve, once $\amlt$ has been suitably
normalized using the Sun (as in Fig.~\ref{fig:vanfig3}).

However, Fig.~\ref{fig:vanfig4} is itself problematic.  Because the KS66
$\fttau$ relation is based on an empirical solar atmosphere, one would expect
that the solid and dotted curves should be nearly coincident, especially at
luminosities and temperatures close to that of the Sun.  That they are not in
particularly good agreement is presumably a consequence of the fact that
modern 1-D, plane-parallel model atmospheres employing the local mixing-length
theory of convection are unable to reproduce the actual $T$--$\tau$ structure
of the Sun adequately.  For instance, \citet{bls95} have shown that the limb
darkening of a theoretical flux-constant solar model does not fit the
observations (as compared with the scaled solar model by \citealt{hm74})
very well (though, interestingly, 3-D models of the solar atmosphere are more
successful; see \citealt{ant99}).

To ensure that there is no significant dependence of the results on the
particular empirical solar atmosphere that is assumed, the dotted curve was
recomputed using the HM74 model as represented by \citet{vp89}.  This solar
atmosphere is preferable to that given by KS66 because the former, but not the
latter, is known to reproduce the solar flux and limb darkening quite well.
[The KS66 $\fttau$ relation was derived solely from a consideration of spectral
line profiles.]  The resultant track (not shown here) agreed with that based on
KS66 atmospheres all the way from the ZAMS to the highest luminosity plotted in
Fig.~\ref{fig:vanfig3} to within $\approx 0.0008$ in $\log\,\teff$ (or $\approx
10$ K).  Of course, to compensate for this small offset in temperature, some
adjustment in the assumed value of $\amlt$ (to 1.96) would be needed 
to obtain a Standard Solar Model with an HM74 atmosphere.  (The required small
decrease in the value of the mixing-length parameter results in the giant-branch
segment of the solar track for the HM74 case being $\approx 20$ K cooler at
$M_{\rm bol} = 2.2$ than the dotted curve.)

Differences in the predicted pressures at $T = \teff$ are the main cause of the
separations (in the horizontal direction) between the three tracks presented in
Fig.~\ref{fig:vanfig4}.  As shown in Figure~\ref{fig:vanfig5}, which plots the
structures of the sub-photospheric layers (down to $\tau\sim 100$) in Standard
Solar Models having MARCS, gray, or KS66 atmospheres, the photospheric pressures
and the variations of $T$ and $P$ with depth differ markedly in the three cases
considered.  (All models must replicate the observed temperature of the Sun at
the photosphere, defined here to coincide with zero depth.)  The best estimate
of the pressure at $T=\teff$ is arguably that indicated by the open circle since
it has been derived from an empirical solar atmosphere.  [Furthermore, because
the calibration of the mixing-length parameter should be based on the most
realistic solar model that can be computed, the preferred estimate of $\amlt$
($\approx 2.0$, assuming the Asplund estimate of $Z_\odot$) is obtained
when a reliable empirical solar atmosphere is used in the computation of a
Standard Solar Model instead of, for instance, a gray or a MARCS atmosphere.]
Unfortunately, it is not known how the physics in MARCS atmospheres should be
modified so that they predict suitably reduced pressures at the photosphere.
Regardless of how it is achieved, a significant reduction in the photospheric
pressure that is predicted by the MARCS solar atmosphere is needed to achieve
consistency with models for the Sun having KS66 (or HM74) atmospheres.

While there is no reason to expect that the same adjustment to the photospheric
pressure should be made at all effective temperatures and gravities, such an
assumption has the intriguing consequence that the resultant evolutionary
sequences are a close match to those obtained when the KS66 $\fttau$ relation
is used to derive the boundary pressure.  This is illustrated in
Figure~\ref{fig:vanfig6}, which compares a solar track that assumes KS66
atmospheres and $\amlt = 2.0$ (the dotted curve) with three different tracks
that employ MARCS atmospheres.  The solid and dashed curves assume that
$\amlt = 1.80$ and 2.00, respectively, while the dot-dashed curve is otherwise
identical to the dashed curve except that the predicted pressures at $T=\teff$
are reduced by $\delta\,\log\,P = 0.1375$ (i.e., the difference between the
open and filled circles in Fig.~\ref{fig:vanfig5}).  It is quite remarkable
that the last of these cases reproduces the KS66 track so well, despite the
variations in the $T$--$\tau$ structures that must exist as a function of
$\teff$ and gravity.  Such variations apparently do not have important
consequences for the photospheric pressure in a systematic sense.

It was, in fact, not necessary to have completely consistent atmosphere-interior
models insofar as the assumed value of the mixing-length parameter is concerned.
As shown in Figure~\ref{fig:vanfig7}, which plots the $T$--depth and $P$--depth
profiles predicted by MARCS atmospheres for the sub-photospheric layers in
models for solar parameters and $\amlt = 1.50$, 1.80, and 1.92, the predicted
photospheric pressure is independent of $\amlt$.   Moreover, since nearly
the same evolutionary tracks are obtained when MARCS atmospheres are attached
to interior models at the photosphere or at $\tau = 100$ (as shown in
Fig.~\ref{fig:vanfig3}), there is little to be gained by making this attachment
at depth (provided that the atmosphere and interior codes incorporate very
similar physics: some counter-examples are discussed below).  The key ingredient
to be obtained from these particular model atmospheres is the photospheric
pressure.

If the loci in Fig.~\ref{fig:vanfig5} for the various Standard Solar Models are
extended to $0.99 {{\cal M}_\odot}$ by integrating the stellar structure
equations, and the resultant $\log T$ and $\log P$ values are plotted as a
function of $\log r$ (instead of depth), one obtains Figure~\ref{fig:vanfig8}.
The inner atmosphere of the Sun (i.e., between the photosphere and the point
where $\tau = 100$) is clearly a very thin layer.  Indeed, because a depth of
$\sim 16\times 10^6$ cm below the photosphere (see Fig.~\ref{fig:vanfig5})
represents only $\sim 0.02$\% of the radius of the Sun, the relatively large
differences between the various loci plotted in Fig.~\ref{fig:vanfig5} are not
discernible in Fig.~\ref{fig:vanfig8}.  Importantly, all of the integrations
yield essentially the same values of temperature, pressure, and radius at $0.99
\msol$ (and, in fact, nearly identical radial variations of $T$ and $P$).

To illustrate the dependence of the integrations down to $0.99 \msol$
on the adopted value of $\amlt$, the case represented by the filled circle,
and assuming the photospheric pressure from the MARCS model atmosphere, was
repeated using $\amlt = 1.71$ instead of 1.80 (the value required by a Standard
Solar Model).  This resulted in the dotted curves, which are actually quite
close to the others that have been plotted, and the values of $\log T$,
$\log P$, and $\log r$ at $0.99 \msol$ indicated in Fig.~\ref{fig:vanfig8} by
the filled squares.  Because the latter are significantly displaced from the
filled circles, the resultant boundary conditions are also quite different,
resulting in a solar model that is $\sim 40$ K cooler than the Sun.  (This
offset is in the expected direction when the assumed value of $\amlt$ is less
than that needed to satisfy the solar constraint; see Fig.~\ref{fig:vanfig4}).

\subsubsection{The Effects of Macroturbulence in Model Atmospheres}
\label{subsubsec:turb}

The MARCS model atmospheres considered thus far did not take into account one
component to the pressure that is normally included; namely, the pressure
arising from macroturbulence, which is due to relatively large scale motions 
in the stellar atmosphere.  (This was disregarded to improve the consistency
with the interior models produced by the Victoria code, since the latter does
not treat this additional physics.)  In the MARCS code, the turbulent pressure
is calculated from $P_{\rm turb} = 0.5\,\rho\,v^2$, where $v$ is the 
characteristic velocity and $\rho$ is the density (for some discussion of this
equation, see Henyey et al.~1965).  The model atmospheres computed for this
study assume that $v = (6.5 - \log g)$ km/s.  (The inclusion of $P_{\rm turb}$
with such depth-independent values of $v$ corresponds closely to a shift in the
surface gravities of models with no macroturbulence, cf.~\citealt{gbe75}:
matters are more complicated if $v$ is assumed to vary with depth.)

However, it seems to be the case that taking a turbulent component to the
pressure into account has no more than a small effect on the predicted $\teff$
scale.  Figure~\ref{fig:vanfig9} compares the structures of the sub-photospheric
layers given by MARCS model atmospheres for solar parameters ($\log g = 4.44$,
$\teff = 5777$ K, and Asplund abundances), computed with and without the
inclusion of turbulent pressure but assuming the same value of the
mixing-length parameter ($\amlt = 1.77$, which is needed to produce a Standard
Solar Model when the turbulent pressure is included in the atmosphere).
The differences in the various quantities are appreciable, though not
large enough to have a big impact on stellar evolutionary models.  As shown in
Figure~\ref{fig:vanfig10}, the tracks that are obtained when turbulent MARCS
atmospheres are used as boundary conditions differ only slightly from those
employing non-turbulent atmospheres.  (At $M_{\rm bol} = 2.3$, the dotted and
dot-dashed curves differ in temperature by $\approx 30$ K.)  This similarity is
due, in part, to the fact that the former calculations assume a somewhat smaller
value of $\amlt$ than the latter (in order to reproduce the properties of the
Sun at the solar age), which thereby compensates for some of the effects due to
the difference in physics. 

There is an additional parameter (besides those associated with the
mixing-length treatment of convection) that can be varied in the model
atmosphere code; namely, the microturbulence, which affects the spectral line
blanketing.  However, variations in this quantity have only a very minor effect
on the predicted temperatures of stellar models.  If, for instance, the assumed
microturbulent velocity is changed from 1 km/s to 2 km/s, the predicted
photospheric pressure of the Sun (or of a model on the lower RGB) decreases by
$\delta\log\,P = 0.007$--0.008.  This change in the boundary pressure results
in stellar models that have lower values of $\teff$ by only a few Kelvin.

\subsection{Solar Abundance Models, Assuming $Z_\odot = 0.0165$ (GS98)}
\label{subsec:gs98}

If the calculation of a solar evolutionary track is repeated assuming the GS98
heavy-element mixture and $Y=0.26764$ (see Table~\ref{tab:tab2}), which implies
$Z_\odot = 0.01651$, together with $\amlt = 1.84$ and boundary conditions based
on fully consistent MARCS model atmospheres, the result is the dashed
curve in Figure~\ref{fig:vanfig11}.  This may be compared with the solid curve,
which assumes the Asplund abundances for several of the elements heavier than
helium (see Table~\ref{tab:tab1}) and the adopted values of the relevant
parameters specified in Table~\ref{tab:tab2}.  (This track is identical to those
plotted as solid curves in Figs.~\ref{fig:vanfig3}, \ref{fig:vanfig4}, and
\ref{fig:vanfig10}.)  The third track that appears in Fig.~\ref{fig:vanfig11}
(the dotted curve) has been taken from the extensive grids of evolutionary
models published by VandenBerg et al.~(2006).  In these calculations, the solar
parameters were taken to be $Y_\odot = 0.2768$, $Z_\odot = 0.0188$ (assuming the
mix of metals given by \citealt{gn93}), and $\amlt = 1.90$.  Moreover, the
pressure at $T=\teff$ was determined by integrating the hydrostatic equation in
tandem with the KS66 $\fttau$ description of the outer atmospheric layers.

As regards the three tracks in Fig.~\ref{fig:vanfig11}: the relative locations
of the subgiant and giant branches are the expected consequences of the 
differences in the CNO abundances (see \citealt{rc85}), which are mainly
responsible for the differences in $Z$, and in the adopted values of $\amlt$,
respectively.  Except for the RGB, the tracks are sufficiently similar that it
is not possible to discriminate between them empirically.  Indeed, the main
conclusion to be drawn from this plot is that the predicted temperatures of
solar abundance giants depend sensitively on the adopted value of $Z_\odot$
(and the detailed heavy-element mixture), which is currently a subject of
considerable controversy (see, e.g., \citealt{ags05}, \citealt{vge07}).

Whether or not the same value of the mixing-length parameter is appropriate for
solar abundance giants as for the Sun must therefore depend on the value of
$Z_\odot$.  \citet{vc03} found that the effective temperatures of M$\,$67
giants, as determined using empirical $(V-K)$--$\teff$ relations, agreed very
well with those predicted by a 4.0 Gyr, $Z = 0.0173$ isochrone that assumed KS66
atmospheric structures.  (In that study, the adopted value of $Z_\odot$ was
0.0188 and M$\,$67 was assumed to have [Fe/H] $=-0.04$, in which case, the
corresponding value of $Z$ is 0.0173.)  Because the prediction of stellar
temperatures is so central to this investigation, it is worthwhile to revisit
this constraint.

\subsection{The Temperatures of M$\,$67 Giants and Their Implications for the
Predicted $\teff$ Scale}
\label{subsec:m67rgb}

If we adopt $E(B-V) = 0.038$ and $(m-M)_V = 9.70$, together with $E(V-K) =
2.78\,E(B-V)$, $A_V = 3.12\,E(B-V)$ and $A_K = 0.34\,E(B-V)$ (\citealt{bb88}),
it is a trivial task to obtain $M_V$ and $(V-K)_0$ values for the giants in
M$\,$67 for which $VK$ photometry has been reported by \citet{hfc92}.  (The
assumed reddening and distance modulus are believed to be current best
estimates; see the discussion of cluster parameters provided by
\citealt{vge07}.)  The temperatures then follow from the empirical relations
between $(V-K)$ and $\teff$ given by \citet{bcp98} and \citet{vb99}, which are
based on the latest determinations of stellar radii from lunar occultations and
Michelson interferometry.  The resultant $(\log\teff,\,M_V)$-diagram is shown in
the left-hand panel of Figure~\ref{fig:vanfig12}.  Note that, at the $M_V$ value
of each star, the temperatures derived from the Bessell et al.~and van Belle et
al.~color transformations are plotted as open and closed circles, respectively. 

The dotted curve is the RGB segment of the same 4.0 Gyr, $Z=0.0173$ isochrone
that provided a very good fit to the M$\,$67 CMD in the \citet{vc03} study.  It
matches the distribution of cluster giants quite well, particularly if compared
with the luminosity and temperature data indicated by the filled circles.  (The
small group of stars at $\log\teff \approx 3.675$ and $M_V \approx 0.9$ plays
no role in this comparison because they are almost certainly horizontal-branch
stars.)  However, this constraint is not nearly as stringent as one would like.
Although the temperatures of many of the stars used in current calibrations of
$(V-K)$ versus $\teff$ have been determined to better than $\sim 1.3$\%, the
standard deviation of the measured values of $\teff$ at a given $V-K$ color is
typically 7\% (van Belle et al.~1999).  Such large uncertainties clearly permit
considerable leeway in the models.

The solid and dashed curves represent, in turn, the giant-branch extensions of
4.2 Gyr, $Z=0.01247$ (Asplund) and 3.9 Gyr, $Z=0.01651$ (GS98) isochrones that
were compared with the \citet{mmj93} CMD of M$\,$67 by \citet{vge07}.  [The
models used in that study are completely consistent with those presented here;
indeed, the tracks for $1.0 \msol$ are identical.  Interestingly, only the
high-$Z$ isochrone predicts a gap where one is observed in $M\,$67 --- which is
potentially a problem for the Asplund estimate of $Z_\odot$, since this open
cluster is known to have very close to solar metal abundances (e.g., see the
results of high-resolution spectroscopy reported by \citealt{ht91};
\citealt{tet00}; \citealt{rsp06}).  However, as noted by VandenBerg et al.,
isochrones for the Asplund metallicity {\it may} provide a good match to the
M$\,$67 CMD in the vicinity of the turnoff if diffusive processes are treated:
further work is needed to investigate this possibility.  Aside from the turnoff,
the 4.2 and 3.9 Gyr isochrones for the Asplund and GS98 metallicities provide
equally satisfactory fits to the cluster main sequence and subgiant branch; see
the VandenBerg et al.~paper.]

There is little doubt that the predicted temperatures along these loci are too
cool (which could be taken as evidence that $\amlt$ is larger in solar abundance
giants than in main-sequence dwarfs;\footnote{Indeed, the 2-D hydrodynamical
simulations of \citet{lfs99} suggest that the proper mixing-length to adopt for
giants and subgiants is $\sim 0.1$ greater than that for solar-type stars.
Following VandenBerg (1991), we estimate that such a change would bring the
model RGB that employs MARCS atmospheres about half-way to the observed sequence
in Fig.~\ref{fig:vanfig12}.  More realistic 3-D models, already available
today (\citealt{cat07}), should be used to further illuminate this possibility.}
however, see below).  The extension of the evolutionary sequences and isochrones
to $M_V \sim -1$ required the calculation of complementary grids of model
atmospheres for $4000 \le\teff\le 5500$ K and $0.0 \le\log g \le 2.5$ in steps
of 250 K and 0.5 dex, respectively.  [Note that, as shown in the right-hand
panel of Fig.~\ref{fig:vanfig12}, the discrepancies found by \citealt{vge07}
between the isochrones represented by the solid and dashed curves and the
faintest cluster giants on the [$(B-V)_0,\,M_V$]-plane persist to higher
luminosities.  Moreover, the close similarity of the comparisons between theory
and observations presented in both panels of this figure indicates that the
adopted $(B-V)$--$\teff$ relations (from \citealt{vc03}) are consistent with
the empirical $(V-K)$--$\teff$ relationships.]

A set of spherical model atmospheres were also computed (for temperatures from
4000 K to 4750 K and $1.0 \le\log g \le 2.5$) to enable us to determine whether
their use as boundary conditions resulted in tracks for upper-RGB stars that
differed in any significant way from those employing plane-parallel atmospheres.
The differences between the models that incorporated the two different types of
model atmospheres were found to be barely discernible, which demonstrates that
sphericity effects are very small and of little consequence for the evolutionary
calculations presented in this investigation.  The dot-dashed curve in
Fig.~\ref{fig:vanfig12} is also based on MARCS model atmospheres, and it is able
to reproduce the properties of M$\,$67 giants (and the Sun) just as well as the
dotted curve (on the assumption of a constant value of $\amlt$).  The SDC
variant of MARCS models, upon which the dot-dashed RGB is based, is introduced
in the next section. 
%

\subsection{Using MARCS SDC Atmospheres as Boundary Conditions}
\label{subsec:sbcatm}

Since, as already noted, the MARCS solar atmosphere does not reproduce the
observed flux and limb darkening of the Sun as well as the HM74 solar
atmosphere, it could be argued that HM74 models, when scaled to the appropriate
$\teff$, provide the best representations of solar-type stars (provided that
the pressure is obtained by integrating the hydrostatic equation at the relevant
surface gravity).  However, MARCS models have the important advantage that
changes in the opacity distributions across the spectrum, and with depth, on
the radiative energy transfer are handled properly by the MARCS code, at least
within the adopted LTE and 1-D assumptions.  These opacity variations occur
systematically with $\teff$, surface gravity, and chemical composition;
consequently, these differential changes should be taken into account if models
with different fundamental parameters are related --- as they are along our
evolutionary tracks.

Thus, rather than adopting theoretical MARCS models directly, it may be 
preferable to use them in a strictly differential sense to correct the HM74
atmospheres for the effects of changes in the basic stellar parameters.  Such
``scaled solar, differentially corrected" (SDC) models would have temperature
structures given by 
$$T(\tau) = \biggl({\teff\over {T_{{\rm eff},\,\odot}}}\biggr)\,{T_{\rm
 HM74}(\tau)} + {T_{\rm MARCS}(\tau)} - \biggl({\teff\over {T_{{\rm
 eff},\,\odot}}}\biggr)\,{T_{{\rm MARCS},\,\odot}(\tau)}\ \ .$$
According to this equation, the HM74 solar atmosphere is first scaled to the
correct effective temperature, and then a correction is applied given by the
difference between the MARCS model at the relevant parameters and the
corresponding scaled solar MARCS model.  This equation is obviously equivalent
to
$$T(\tau) = T_{\rm MARCS}(\tau) + \biggl({\teff\over {T_{{\rm
  eff},\,\odot}}}\biggr)\,[T_{\rm HM74}(\tau) - T_{{\rm
  MARCS},\,\odot}(\tau)]\ \ ,$$
i.e., the adopted $\fttau$ relation is obtained by correcting the relevant
MARCS model by the {\it scaled} difference between the HM74 and MARCS solar
atmospheres.  All scaling is done with respect to the Rosseland mean optical
depth: with the resultant $T$--$\tau$ structure so obtained, the equation of
hydrostatic equilibrium is then integrated on the assumption of the relevant
chemical composition and surface gravity to obtain the pressure as a function
of depth in the atmosphere.  Note that, for a model having solar parameters,
the above equation reduces to $T(\tau) = T_{\rm HM74}(\tau)$.  (As already
mentioned, it is preferable to use the HM74 atmosphere in this exercise,
instead of that by KS66, because it has been much more thoroughly studied
and found to provide good consistency with the observed flux, limb darkening,
and spectral line profiles of the Sun.)

Because the HM74 solar atmosphere is not defined at optical depths greater
than $\tau = 10$, where the uncertainties will already be large because so
little of the observed light originates at such depths, we have opted to fit
the MARCS SDC atmospheres to the interior models at $\tau = 5$.  With just
this change to the treatment of the atmospheric layers and the assumption
of $\amlt = 2.01$, which is required to satisfy the solar constraint in this
case, evolutionary tracks that are otherwise identical to those plotted as
solid curves in Fig.~\ref{fig:vanfig13} were computed, along with an isochrone
for 4.2 Gyr.  Taken at face value, the giant-branch segment of this isochrone
(the dot-dashed curve in Fig.~\ref{fig:vanfig12}) does the best job of all the
cases that have been plotted in reproducing the observed RGB of M$\,$67 (though
the uncertainties in the data are too large to differentiate between the dotted
and dot-dashed curves).

On the $[(B-V)_0,\,M_V]$-plane (see the right-hand panel of
Fig.~\ref{fig:vanfig12}), the lower RGB segment of the dot-dashed isochrone is
approximately mid-way between the dashed curve and the observed giant branch.
(There is no obvious discrepany in the left-hand panel, though only two stars
fainter than $M_V \sim 2$ have measured temperatures.)  However, a difference
$\approx 0.03$--0.05 mag in $B-V$ at $2 \lta M_V \lta 3.5$, is within the
uncertainties of current color--$\teff$ relations.  For instance, the
transformations reported by \citet{hbs00} yield a value of $B-V = 1.023$ for a
star having [Fe/H] $=0.0$, $\teff = 4750$ K, and $\log g = 3.0$ (which is
appropriate for a lower RGB star in M$\,$67), while $B-V = 1.050$ is obtained
from the $(B-V)$--$\teff$ relations reported by \citet{cas99}, on the assumption
of the same stellar parameters.  [For the reasons discussed by \citet{vc03},
the Castelli transformations were considered to be more realistic, but that was
a judgment call and not necessarily the right one (particularly if the Asplund
value of $Z_\odot$ is correct).  At the same time, it should be appreciated that
model-atmosphere-based synthetic colors generally do not agree well with those
given by empirical color--$\teff$ relations (e.g., \citealt{sf00}), especially
for cool stars.  Consequently, it is not possible to obtain anywhere near as
satisfactory a fit of isochrones to the entire CMD of M$\,$67 as that presented
by VandenBerg et al.~(2006) using, for instance, either the Houdashelt et al.~or
the Castelli transformations.  This is the reason why VandenBerg \& Clem relied
on empirical color--$\teff$ relationships whenever it was possible to do so.]
In any case, if the MARCS SDC atmospheres are assumed to be realistic boundary
conditions, one may concluded that, at least for solar abundance stars, $\amlt$
does not vary significantly (if at all) with changes in $\teff$ between 4000 K
and 7000 K or in $\log g$ values from 0.0 to 5.0.

Given that the effect of using SDC atmospheres for the Asplund metallicity as
boundary conditions for stellar models is to move the predicted RGB from the
location of the solid curve to that of the dot-dashed curve (see
Fig.~\ref{fig:vanfig12}), one might anticipate that the use of SDC atmospheres 
for the GS98 abundances would imply a giant branch that is somewhat too
hot/blue.  This assumes that the separation, in the horizontal direction,
between the latter and the dashed curve would be comparable to that between the
dot-dashed and solid curves.  We have not examined this possibility, mainly
because the difference in $\teff$ between the dashed and solid curves is
comparable to the uncertainty in the empirically derived temperatures (at 
least $\pm 50$--100 K).  Thus, even if the predicted RGB for the GS98 case
turned out to be $\sim 75$ K hotter than the dot-dashed curve, it would still
be within the $1\,\sigma$ uncertainty of the measured temperatures of M$\,$67
giants, and it would be incorrect to conclude, for instance, that models for
giants must assume a smaller value of $\amlt$ than those for dwarf stars.  The
observational and theoretical uncertainties are such that a small variation in
$\amlt$ with evolutionary state cannot be ruled out.

\section{Models for [Fe/H] $= -2.0$ Stars}
\label{sec:mpoor}

To investigate the impact of using different treatments of the atmosphere on
the predicted $\teff$ scale at low $Z$, evolutionary tracks have been computed
for [Fe/H] $= -2.0$, on the assumption of the mix of heavy elements listed in
Table~\ref{tab:tab3}.  To obtain this particular mixture, the Asplund $\log N$
abundances for the Sun (Table~\ref{tab:tab2}) were reduced by 2.0 dex and then
adjusted appropriately so that the resultant [m/H] values (see the fourth
column in Table~\ref{tab:tab3}) were close to the measured values in [Fe/H]
$\approx -2$ stars (see \citealt{cds04}).  Note that most of the $\alpha$
elements are enhanced by 0.3 dex (oxygen by 0.5 dex), while a few elements (Na,
Cr, and Mn) are underabundant relative to a scaled-solar mixture. As far as
helium is concerned, $\log N_{\rm He} = 10.92$ was assumed, so that the
resultant mass-fraction abundances came out to be $X = 0.75149$, $Y =
0.248233$ (in agreement with the value implied by the concordance between WMAP
and Big-Bang nucleosynthesis; see \citealt{cvd04}), and $Z = 2.8\times 10^{-4}$.
Opacities and a set of MARCS model atmospheres for $4000 \le\teff\le 8000$ K and
$0.0 \le\log g\le 5.0$ were computed for these chemical abundances.  Moreover,
$\amlt = 1.80$ was adopted, in both the atmosphere and interior models, in order
to be consistent with the requirements of a Standard Solar Model.  (Stellar
models employing metal-poor MARCS SDC atmospheres are discussed below.)

It is well known (e.g., see \citealt{gbe75}; \citealt{ir85}) that convection
extends to shallower (i.e., smaller) optical depths in a metal-poor atmosphere
than in a metal-rich atmosphere having a similar $\teff$ and gravity.  This
raises the concern that models for metal-deficient stars may be much more 
dependent on where the atmospheres are attached to the interior structures
(i.e., at the photosphere or at some deeper layer) than those for solar
abundances. Indeed, Runge-Kutta integrations of the stellar structure equations,
beginning at the photosphere, are not able to reproduce the predictions of
metal-deficient model atmospheres for the deeper layers particularly well.
This is shown in Figure~\ref{fig:vanfig13}, which compares the variations of
several quantities in the sub-photospheric layers, as predicted by MARCS
atmospheres for a star having $\teff = 6398$ K, $\log g = 4.59$, and the metal
abundances described above, with the results of Runge-Kutta integrations
performed by the Victoria evolutionary code.  [A similar comparison (not shown)
was carried out for a lower RGB model having $\teff = 5444$ K and $\log g =
3.35$: it looked qualitatively the same as Fig.~\ref{fig:vanfig13}.]   As noted
previously, in connection with Fig.~\ref{fig:vanfig1}, which shows that much
closer agreement is obtained in the case of a solar model, the differences
between the solid and dashed curves are consistent with the expected
consequences of the diffusion approximation.

However, it turns out that these differences do not have a significant effect
on the predicted effective temperatures of stellar models, probably because the
atmospheres are so thin that slight variations in the atmospheric structures
give rise to no more than minor perturbations to the boundary conditions.  
Figure~\ref{fig:vanfig14} shows that the tracks which are obtained for a $0.8
\msol$ star when MARCS model atmospheres are attached at the photosphere or at
$\tau = 100$ are essentially indistinguishable (compare the solid and dashed
curves).  Also plotted in this figure are evolutionary sequences for the same
mass and chemical abundances, but assuming that the outer atmospheric
temperature distributions are given by KS66 or gray $\fttau$ relations.
These calculations, which also assume the values of $\amlt$ from the 
appropriately calibrated solar models (see Fig.~\ref{fig:vanfig3}), are both
remarkably close to the tracks utilizing MARCS model atmospheres.  Although 
one might be tempted to conclude from this that stellar models for very 
metal-poor stars do not have a very sensitive dependence on the treatment of
the atmosphere, it must be kept in mind that the effects of differences in the
assumed mixing-length parameters have conspired to produce the apparent
similarity of the tracks that have been plotted.

Indeed, the small differences that exist are contrary to expectations.  In
particular, it seems odd that the main-sequence portions of the tracks that
employ MARCS model atmospheres would be warmer than those using gray atmospheric
structures, and that the gray and KS66 models would be nearly coincident,
despite being based on very different $\fttau$ relations.  These anomalies have
arisen because the respective values of $\amlt$ that have been assumed are
quite different.  For instance, suppose that the models with gray atmospheres
are recomputed on the assumption that $\amlt = 2.0$; i.e., the same
value of the mixing-length parameter that is needed to produce a Standard Solar
Model when the KS66 atmospheres are assumed.  If the models with gray
atmospheres are revised in this way, a markedly different comparison
between the different tracks is obtained, as shown in Figure~\ref{fig:vanfig15}.

Now the models with the gray atmospheres are appreciably hotter than those 
that describe the atmospheric layers using the KS66 $\fttau$ relation, and the
tracks using MARCS atmospheres lie between them.  This certainly corresponds
more closely to one's (possibly naive) expectations.  However, a significantly
warmer giant branch is not predicted by models that use MARCS SDC atmospheres
(for [Fe/H] $=-2.0$) as boundary conditions.  As shown in
Fig.~\ref{fig:vanfig16}, evolutionary sequences based on SDC atmospheres hardly
differ from those employing KS66 or HM74 atmospheres.  This is especially true
at low $Z$, but even at the solar metallicity, the differences between these
cases are small.  In other words, models having KS66 or HM74 atmospheres are
able to reproduce the calculations based on MARCS SDC atmospheres rather well,
nearly independently of the assumed metal abundance --- which is a very
surprising result!

Note that, for each of the four cases considered in Fig.~\ref{fig:vanfig16},
$\amlt$ has been properly calibrated using the Sun.  In fact, the differences
in the assumed values of $\amlt$ are largely responsible for the separations of
the RGB loci: not only do the predicted temperatures at a given $M_V$ tend to
increase with increasing $\amlt$ (as expected), but the shift in $\teff$ due
to a given change in $\amlt$ is much larger for metal-rich giants than for
those of low $Z$ (which is also consistent with expectations; see Fig.~3 by
\citealt{van83}).  Had a constant value of the mixing-length parameter been
assumed, the differences between the tracks would have been considerably
smaller (but then, of course, the solar constraint would not have been satisfied
equally well by the different calculations).

Two other points should be made concerning Fig.~\ref{fig:vanfig16}.  First, 
the solar tracks represented by the dashed and dotted curves required slightly
different values of $\amlt$ to satisfy the solar constraint, even though both
describe the upper atmosphere of the Sun using the HM74 $\fttau$ relation.
However, the MARCS SDC atmospheres take macroturbulence into account, and as
already noted in \S~\ref{subsubsec:turb}, a slight reduction in the assumed
value of $\amlt$ is needed to compensate for this additional physics.  Second,
predicted temperatures have a fairly significant dependence on the fitting 
point that is adopted when MARCS SDC atmospheres are attached to interior
models.  This is not entirely unexpected because the temperature structures of
these atmospheres (for the convective, sub-photospheric layers, in particular),
being based on the empirical HM74 model for the Sun, may well be quite different
from those predicted by the models that assume the mixing-length theory.
Indeed, this may explain why the the value of $\amlt$ that is required by a
Standard Solar Model must be changed from 1.94 to 2.01 when the SDC atmospheres
are attached at the photosphere or at $\tau = 5$, respectively.
     
Despite this concern, the sequences of models having KS66 or HM74 atmospheres
are remarkably similar to those using MARCS SDC atmospheres, both at the solar
metallicity and at [Fe/H] $=-2.0$.  Fig.~\ref{fig:vanfig16} has certainly 
reinforced the indications from Fig.~\ref{fig:vanfig6} that the variations in
the predicted $T$--$\tau$ structures from model atmospheres encompassing wide
ranges in the fundamental stellar parameters cannot be very large.  This
conjecture is corroborated by Figure~\ref{fig:vanfig17}, which compares the
variations of $(T/\teff)^4$ with $\log\,\tau$ from MARCS SDC model atmospheres
appropriate to giants and dwarfs having [Fe/H] $= 0.0$ and $-2.0$ with those
given by the HM74 or gray $\fttau$ relations.  The similarity of the different
loci between $\log\,\tau\sim -4$ and $\sim +0.2$ is striking.  Only the gray
atmosphere is offset from the others by a moderately large amount --- and
because it is so discrepant from the predictions of proper model atmospheres,
even for low [Fe/H] values, it seems inadvisable to use such atmospheres in
deriving the surface boundary conditions of stellar models.

On the other hand, it is very encouraging that (i) the predicted $T$--$\tau$
relations from MARCS SDC atmospheres, throughout most of the line-forming
region, are very robust, in the sense that changes to them arising from
variations in the global properties of stars (i.e., $\teff$, surface gravity,
and chemical composition) are small, and (ii) the HM74 solar atmosphere provides
a reasonably good average of the model atmosphere predictions over a large
region of parameter space.  Thus, the commonly used practice of determining
the photospheric pressure by integrating the hydrostatic equation in conjunction
with a scaled-solar $\fttau$ relation (either KS66 or HM74) has much to commend
it.  Indeed, the differences between evolutionary calculations that employ KS66
or HM74 atmospheres and those based on the more sophisticated MARCS SDC model
atmospheres are probably too small to be observationally detectable, given
current uncertainties in the measured properties of stars.  To amplify on this
point, we now turn to some comparisons of our isochrones for [Fe/H] $= -2.0$
with the observed CMD of the globular cluster M$\,$68, which has [Fe/H] $=
-1.99\pm 0.06$, according to the results of high-resolution spectroscopy
carried out by \citet{cg97}.   


\subsection{Fits of Isochrones to the CMD of M$\,$68}
\label{subsec:m68}

The giant and subgiant sequences of M$\,$68 are especially well defined in the
``standard field" photometry provided by \citet{st00};\footnote{See
http://cadcwww.hia.nrc.ca/standards.} consequently, we have opted to use these
data to determine the mean locus describing the evolved stars.  Stetson's
observations and our eye-estimated, hand-drawn curve through them are plotted
in Figure~\ref{fig:vanfig18}.  The solid curve merges smoothly into the fiducial
that \citet{van00} used to represent the stars fainter than $V=19$ in the deep
CMD obtained by \citet{wal94}.  In this way, a CMD has been obtained for M$\,$68
that extends from $\sim 4.5$ mag above the cluster turnoff to $\sim 3.5$ mag
below it.

Grids of evolutionary tracks were computed on the assumption that the pressure
at the photosphere was obtained (i) from fully consistent (standard) MARCS model
atmospheres, which may be fine for metal-deficient stars even though they have
some short-comings in reproducing solar observations, and (ii) by integrating
the equation of hydrostatic equilibrium on the assumption of the KS66 $\fttau$
relation.  (Only these two cases are considered because the low-metallicity
tracks that were computed for the different treatments of the atmosphere are
so similar; see Figs.~\ref{fig:vanfig14} and~\ref{fig:vanfig16}.  Note, in
particular, that the evolutionary sequence which made use of MARCS SDC
atmospheres tends to lie on or between those for the above two cases.)
Needless to say, the calibrated values of $\amlt$ from the respective Standard
Solar Models were adopted.

Isochrones for ages of 12, 14, and 16 Gyr were derived by interpolation in the
tracks, and then they were transposed to the observed plane using the
$(V-I)$--$\teff$ relations given by \citet{vc03}.  If the reddening implied by
the Schlegel et al.~(1998) dust maps is assumed, and the apparent distance
modulus is determined by main-sequence fits to the isochrones, we obtain the
comparisons between theory and observations shown in Figure~\ref{fig:vanfig19}.
The fit to the observations in the upper panel is quite agreeable, aside from
a slight offset between the isochrones and the cluster fiducial at $M_V \lta
0.0$, which suggests that there is a minor problem with the predicted
temperatures or colors for metal-poor giants.  (Note that the indicated age,
$\approx 14.5$ Gyr, would be reduced by $\sim 10$--12\% if diffusive processes
were taken into account; see \citealt{vrm02}.)

Because the ZAMS for the models using KS66 atmospheric structures is cooler 
than that based on MARCS model atmospheres, this case (see the lower panel in
Fig.~\ref{fig:vanfig19}) leads to a somewhat larger distance modulus, a younger
age, and a more pronounced discrepancy along the giant branch.  To demonstrate
the sensitivity of the model fits to a change in the assumed distance, zero-age
horizontal branch (ZAHB) loci were computed on the assumption of the envelope
helium abundance and core mass given by a $0.8 \msol$ model when the He-burning
luminosity (at the tip of the RGB) had surpassed 100$\,L_\odot$.  (A ZAHB model
was calculated according to the standard procedure whereby an initial structure
for a given mass was constructed to have the helium core mass and envelope
chemical abundance profiles from a suitable giant-branch precursor, and then
relaxed over many short time steps until it attained an age of 2 Myr.  The
20--30 models that constitute a ZAHB locus were produced by successively
reducing the envelope mass, while maintaining a fixed core mass.)

If the distance to M$\,$68 is derived by fitting the sample of cluster
horizontal-branch stars observed by \citet{wal94} to this ZAHB, one obtains
$(m-M)_V = 15.26$.\footnote{\citet{vc03} previously found $(m-M)_V = 15.18$
using the ZAHB locus computed by VandenBerg et al.~(2000) for [Fe/H] $= -2.01$
and [$\alpha$/Fe] $=0.3$.  Approximately 0.06 mag of the difference is due to
the assumption, in the present ZAHB models, of a higher helium content and a
slightly larger core mass, which is a consequence of using the conductive
opacities by \citet{pbh99} instead of those by \citet{hl69}.  The remainder can
be attributed to the change made by VandenBerg et al.~(2006) to the VandenBerg
\& Clem bolometric corrections so that $M_{V,\,\odot} = 4.82$ instead of 4.84.
Since the VandenBerg et al.~(2000) ZAHB loci were found to be $\sim 0.02$ mag
fainter than the empirical estimates of RR Lyrae luminosities by \citet{dc99}
and \citet{ccc05}, the ZAHB used in this study to provide the lower bound to
the distribution of HB stars in M$\,$68 is brighter than these determinations
by $\sim 0.06$ mag.}  To identify which isochrone provides the best match to
the observed CMD in the vicinity of the turnoff, it is necessary to adjust the
isochrones to the red by 0.026 mag and 0.013 mag in order that the MARCS and
KS66 atmosphere-interior models for the unevolved stars, respectively, reproduce
the observed ZAMS.  The results of this exercise are shown in
Figure~\ref{fig:vanfig20}: the inferred age of M$\,$68 (which is independent of
the treatment of the atmosphere, as it should be) is close to 12 Gyr.

The main difficulty with this interpretation of the observed CMD is the large
discrepancy between the predicted and observed giant branch loci, which was
obviously exacerabated by the adopted redward color shift. However, this problem
would disappear if the giant-branch segments of the isochrones were hotter by
only $\delta\log\teff = 0.01$.  This is demonstrated by the dashed curves in
Fig.~\ref{fig:vanfig20}, which indicate where the RGBs of the 12 Gyr isochrones
would be located if the model temperatures were increased by this amount
(assuming the same color transformations that were used for the solid curves;
i.e., those by \citealt{vc03}).  Since the giant branches that are plotted in
Fig.~\ref{fig:vanfig15} for the gray and KS66 cases differ in temperature, at a
given luminosity, by only $\delta\log\teff\approx 0.015$, it is easily possible
that the assumed distance in Fig.~\ref{fig:vanfig20} is accurate and that the
giant-branch discrepancies have arisen mainly because the model temperatures
are too low by $\approx 100$ K.


However, there may well be other contributing factors.  For instance, the color
transformations by \citet{vc03}, which we have used here, may be too red for
low gravity stars; though this seems unlikely because they are essentially
identical with those predicted by \citet{bg89}, which satisfy a number of
observational constraints, and because the majority of alternative
color--$\teff$ relations (by, e.g., \citealt{cas99}; \citealt{lcb98}) are even
redder (see the comparison plots provided by VandenBerg \& Clem).  It may also  
be the case that the adopted metallicity of M$\,$68 is too high.  \citet{ki03}
have obtained [Fe/H] $\approx -2.4$ for this system using high-resolution
spectroscopy, but relying on Fe II lines instead of Fe I lines, which were the
basis of the metallicity determinations by \citet{cg97}.  In this regard,
\citet{asp04} has concluded from his 3-D, LTE calculations that abundances
derived from minority species (like Fe I) will be over-estimated by $\gta 0.3$
dex if 1-D model atmospheres (which were used by both Kraft \& Ivans and
Carretta \& Gratton) are employed in the analysis.  According to Asplund, ``it
is clear that globular cluster metallicities should be based on Fe II lines".
As is well known, stellar models for lower metallicities have hotter giant
branches if all other parameters are kept the same.  

If anything, the models for metal-poor main-sequence stars are too hot/blue,
due to the fact that they do not take diffusive processes into account.
However, models with uninhibited diffusion are ruled out by the observed Li
abundances in low-metallicity field turnoff stars (the Spite plateau; see
\citealt{rmr02}) and the smaller-than-expected variation in the abundances of
heavy elements between the turnoff and the lower RGB in NGC$\,$6397
(\citealt{kgr06}).  These constraints can be satisfied if turbulent mixing
below convective envelopes is invoked, which has the additional consequence
that the predicted temperatures at the turnoffs of isochrones, where the
effects of diffusion are most pronounced, do not differ by more than $\approx
60$ K from those given by non-diffusive models (see the plots provided by
\citealt{vrm02}).  It is difficult to say how this result would be
altered if fully consistent model atmospheres were used as boundary conditions.
This should be explored, though such a project would be computationally 
demanding because small grids of model atmospheres would have to be calculated
along each evolutionary track to follow the surface abundance variations
arising from the combined effects of gravitational settling, radiative
accelerations, and turbulent mixing.  

Although our analysis of M$\,$67 giants suggested that $\amlt$ does not vary
with $\teff$ or $\log g$, it is clear from the above discussion that the
many uncertainties at play in the case of the globular cluster M$\,$68 make
it impossible to say whether or not the mixing-length parameter varies either
with [Fe/H] or with evolutionary state within a metal-deficient system.

\section{Conclusions}
\label{sec:conclude}
This investigation has been carried out to examine the consequences for stellar
models of using fully consistent MARCS model atmospheres to describe their outer
layers and to explore how such models differ from those that base the
determination of the photospheric pressure on either the classical gray or KS66
$\fttau$ relationship.  Nearly all large grids of stellar models computed
to date have opted for the latter approach, which must lead to systematic errors
in the predicted $\teff$ scale, since the same $T$--$\tau$ structure (scaled to
the relevant value of $\teff$) cannot apply to stars in all parts of the H-R
diagram.  One would expect that the use of proper model atmospheres to determine
the boundary conditions of stellar models would result in effective temperatures
that are systematically more correct.

Solar abundance models that describe the atmosphere using the KS66 $\fttau$
relation predict considerably warmer giant branches than those using MARCS or
gray atmospheres, and in fact, the application of the latest $(V-K)$--$\teff$
relations to giants in the [Fe/H] $\approx 0.0$ open cluster M$\,$67 suggest
that the hotter temperatures are the most accurate ones.  This study has shown
that the impact of the diffusion approximation for describing the radiative
transfer in stellar interior models is inconsequential (at least for solar-type
stars), insofar as nearly the same tracks are obtained whether MARCS
atmospheres, which obtain ``exact" solutions to the transfer equation, are
attached at the photosphere or at $\tau = 100$.  The higher temperatures that
are obtained when scaled solar model atmospheric structures are assumed arise
because the solar model for this case requires $\amlt = 2.00$, whereas the solar
model with a MARCS atmosphere requires $\amlt = 1.80$.  The failure of current
1-D model atmospheres to reproduce the actual temperature structure of the Sun
is undoubtedly contributing to some (perhaps most) of the differences between
the tracks using MARCS and scaled solar atmospheres as boundary conditions.

Stellar models with gray atmospheres can hardly be expected to be representative
of metal-rich stars like the Sun, nor does it make any sense that such models
for giant stars predict cooler effective temperatures than those using the KS66
$\fttau$ relation to describe the atmospheric layers.  And yet, this is
precisely what happens when the two sets of models are made to satisfy the solar
constraint by setting the value of $\amlt$ appropriately.  (The best estimate
of $\amlt$ is obviously the value obtained when an empirical solar atmosphere
is used in the calculation of a Standard Solar Model, not a gray atmosphere.)
It is perhaps more reasonable to expect that gray atmospheres would be most
relevant for models of very metal-deficient stars, but how realistic is such an
assumption?  To investigate this question, as well as the issues discussed in
the previous paragraph, ``scaled solar, differentially corrected " (SDC)
MARCS atmospheres were constructed that yield the empirical $T$--$\tau$
structure of the Sun (specifically that derived by HM74, which is an improvement
over the earlier KS66 description) if solar parameters are assumed, and yet
retain the differential effects on the $\fttau$ relations that are predicted by
theoretical MARCS atmospheres for different values of $\teff$, $\log g$, and
metallicity.

It is certainly a most interesting and important result of this study that
the temperatures structures of these SDC atmospheres, from $\tau\sim 10^{-4}$
to $\tau\sim 1$, are rather weak functions of the basic stellar parameters, and
that, in the mean, they are well reproduced by the HM74 (or KS66) $\fttau$
relations.  As a result, stellar models that use SDC atmospheres as boundary
conditions satisfy the solar constraint on the assumption of $\amlt\approx 2.0$
(by design) and they reproduce the temperatures of M$\,$67 giants just as well
as the models employing KS66 atmospheres.  Moreover, the differences between the
tracks with these two treatments of the atmospheric layers are in very good
agreement at [Fe/H] $= -2.0$ as well.  (Even at this metallicity, gray 
atmospheres are not good approximations to proper model atmospheres.)

However, it is not necessarily the case that SDC atmospheres represent those of
metal-deficient stars better than standard MARCS models just because the latter
are problematic for metal-rich dwarfs.  Furthermore, while the MARCS models
(including the SDC version) treat the changing character of the radiative energy
transfer as the distribution of opacity varies with fundamental stellar 
parameters, the energy balance may well be qualitatively different in real
stellar atmospheres, as compared with the predictions of mixing-length models,
due to variations in the convective fluxes in the outer layers.  Thus, if the
differential effects of convection for stars of different $\teff$, surface 
gravity, and metallicity are considerable, they may offset or even counteract
the effects calculated by models (such as ours) that are in radiative 
equilibrium in the visible layers.  It would be just fortuitous if, for
instance, those convection effects were such that the scaled solar atmospheres
(i.e., HM74 or KS66) were closest to the true structure.  As far as the choice
between SDC and standard MARCS models is concerned, it seems clear that, for
stars having close to solar parameters, the SDC models are to be preferred.
Fortunately, both SDC and standard MARCS atmospheres imply quite similar $\teff$
scales for metal-poor stars (if $\amlt$ is obtained from the respective
Standard Solar Model).

\acknowledgements
We thank Santi Cassisi for a very thoughtful and helpful report on this paper. 
This work has been supported by the Natural Sciences and Engineering
Research Council of Canada through a Discovery Grant to DAV, and by grants from
the Swedish Research Council to BE, KE, and BG.

\clearpage
\begin{figure}
\plotone{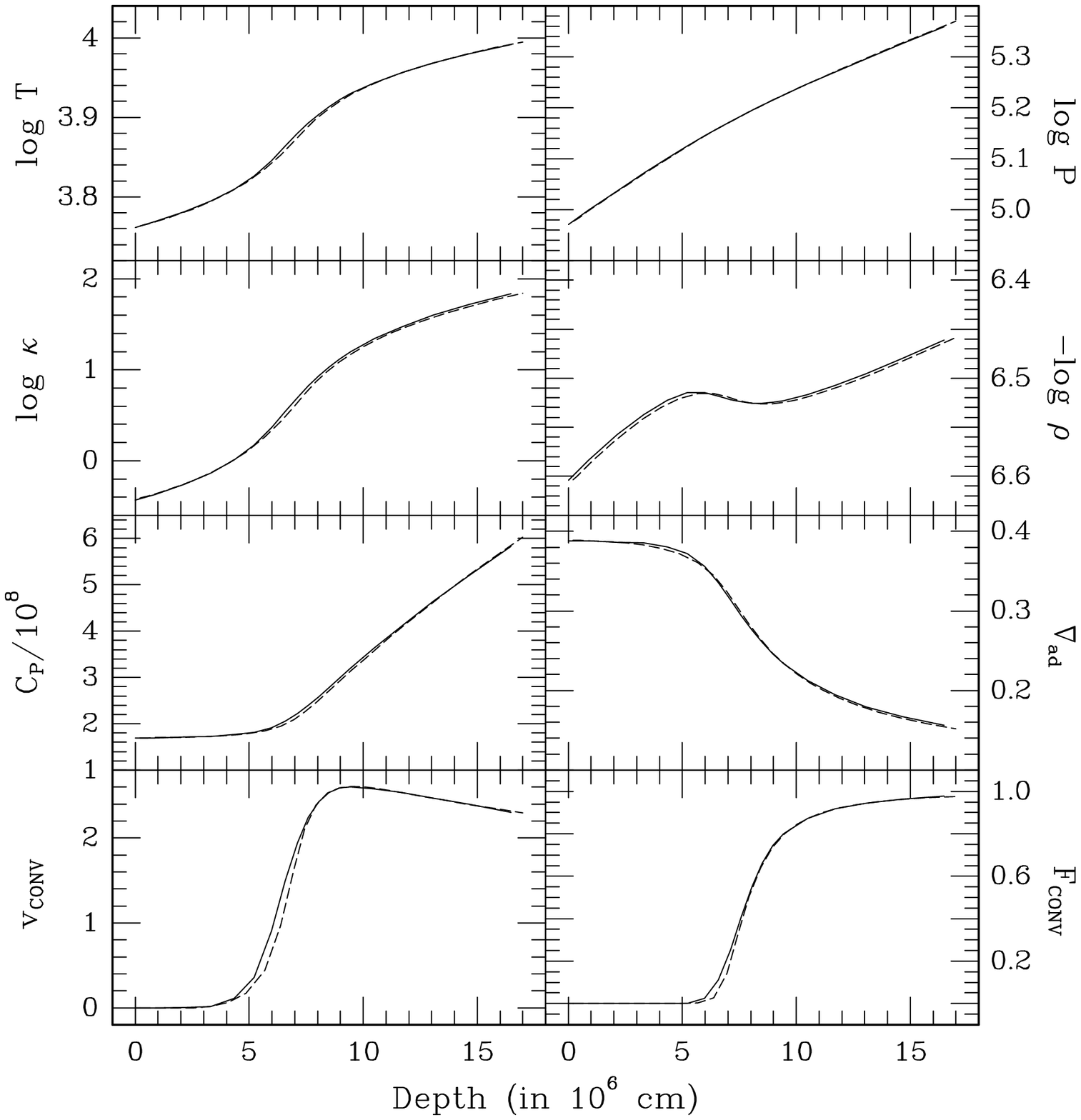}
\caption{Comparison of various thermodynamic quantities ($T$, $P$, $\rho$,
$C_P$, $\nabla_A$), the opacity ($\kappa$), the convective velocity
($v_{\rm conv}$), and the fraction of the total flux that is carried by
convection ($F_{\rm conv}$) in the sub-photospheric layers of a solar model,
as predicted by the MARCS atmosphere and Victoria stellar interior codes (solid
and dashed curves, respectively), on the assumption of the Asplund heavy-element
mixture (see Table~\ref{tab:tab1}).  Both calculations assume $\amlt = 1.80$.}
\label{fig:vanfig1}
\end{figure}

\clearpage
\begin{figure}
\plotone{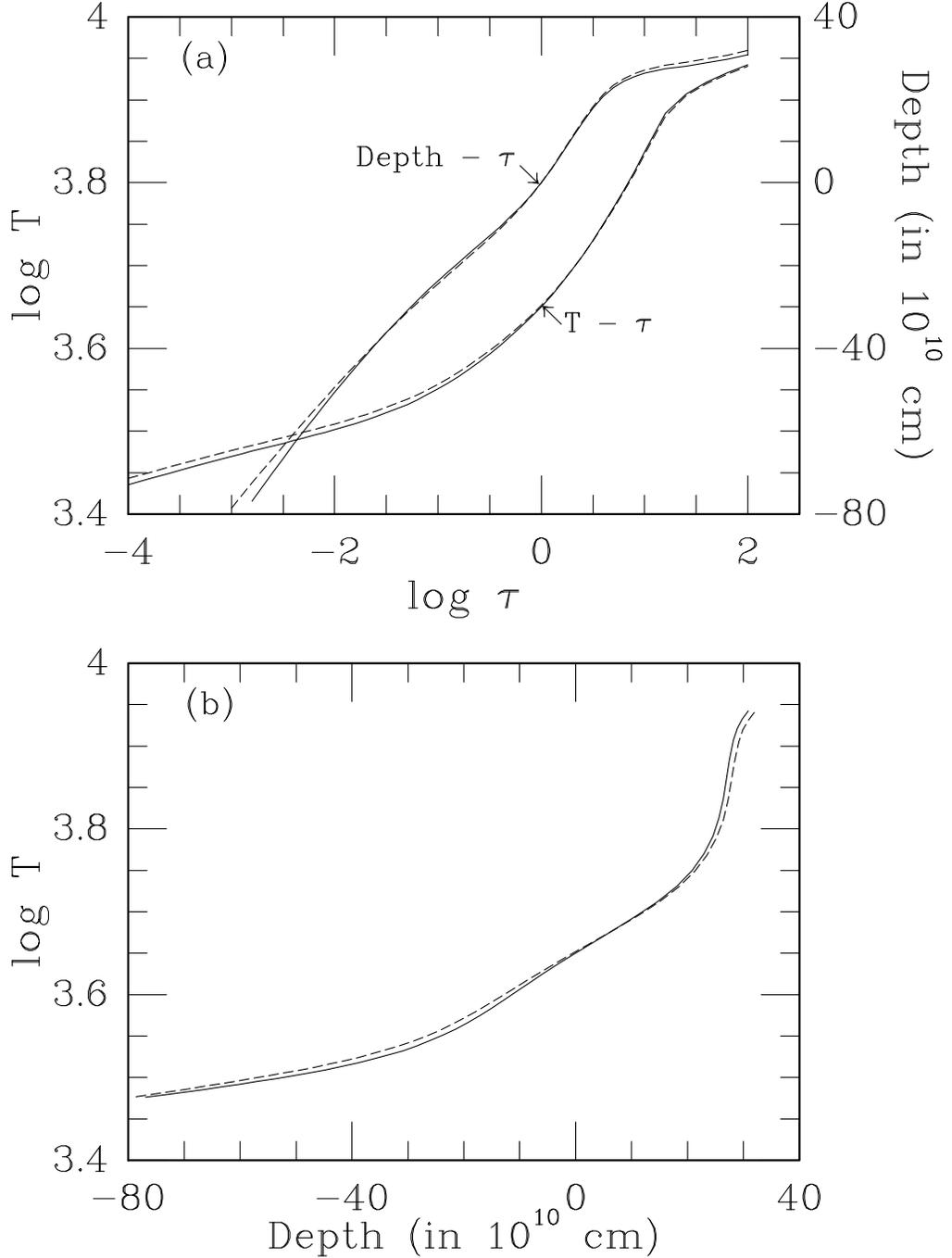}
\caption{The variations of $T$ and geometrical depth with $\tau$ (upper panel)
and of $T$ with depth (lower panel) from spherical model atmospheres for stars
having $\teff = 4000$ K, $\log g = 0.0$, $Z=0.05$, and masses of $0.5 \msol$
(solid curves) and $5.0 \msol$ (dashed curves), respectively.  Note that the
depth zero-point has been arbitrarily chosen to coincide with that layer where
$\tau = 1.0$.}
\label{fig:vanfig2}
\end{figure}

\clearpage
\begin{figure}
\plotone{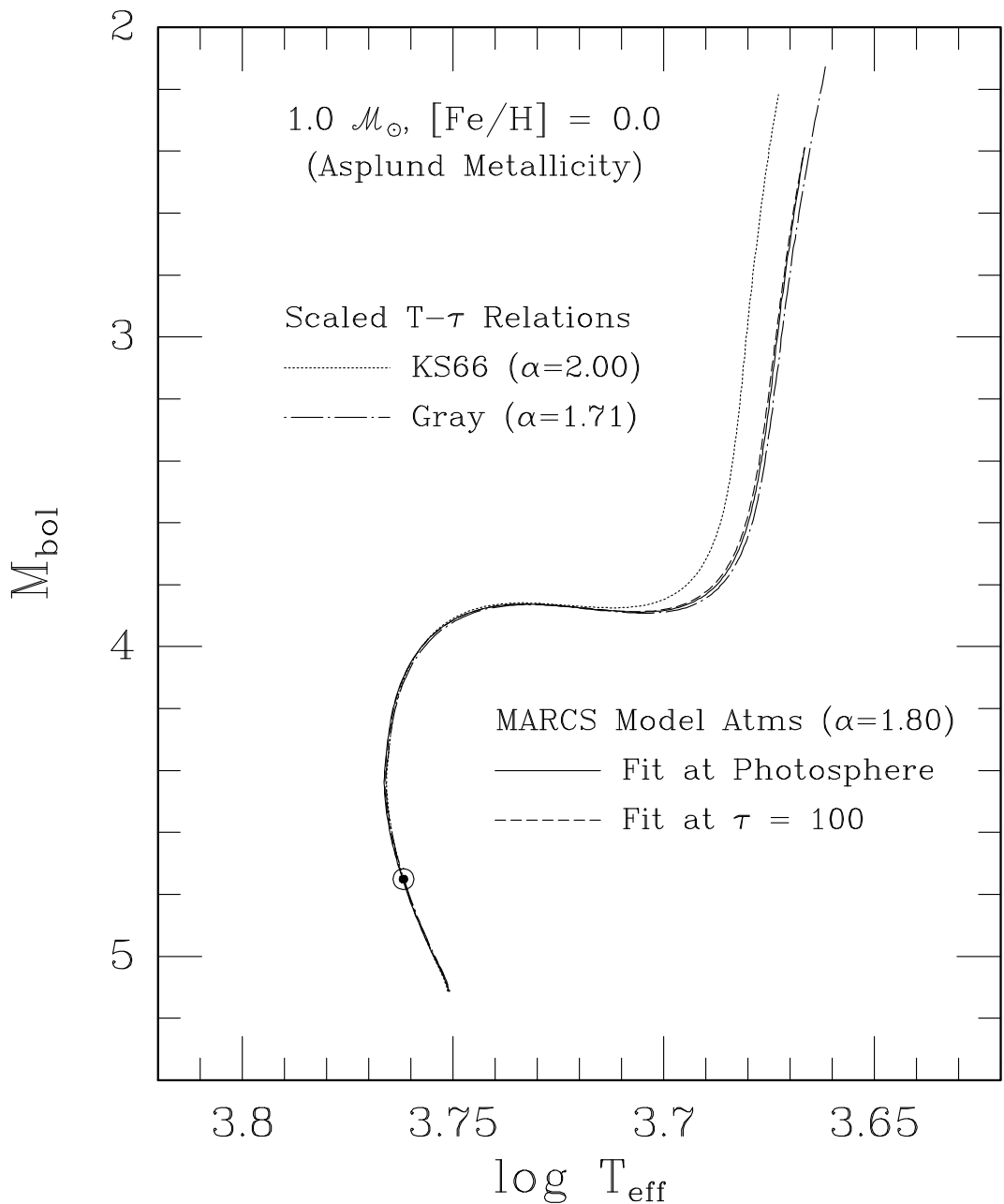}
\caption{Plot on the H-R diagram of evolutionary tracks for $1.0 \msol$ stellar
models having $Y=0.257578$ and the Asplund heavy-element abundances (see
Tables~\ref{tab:tab1} and~\ref{tab:tab2}) with different treatments of the
atmospheric layers, as indicated.  In order for each case to satisfy the solar
constraint (note the location of the solar symbol), the mixing-length parameter
had to be set to the specified values.}
\label{fig:vanfig3}
\end{figure} 

\clearpage
\begin{figure}
\plotone{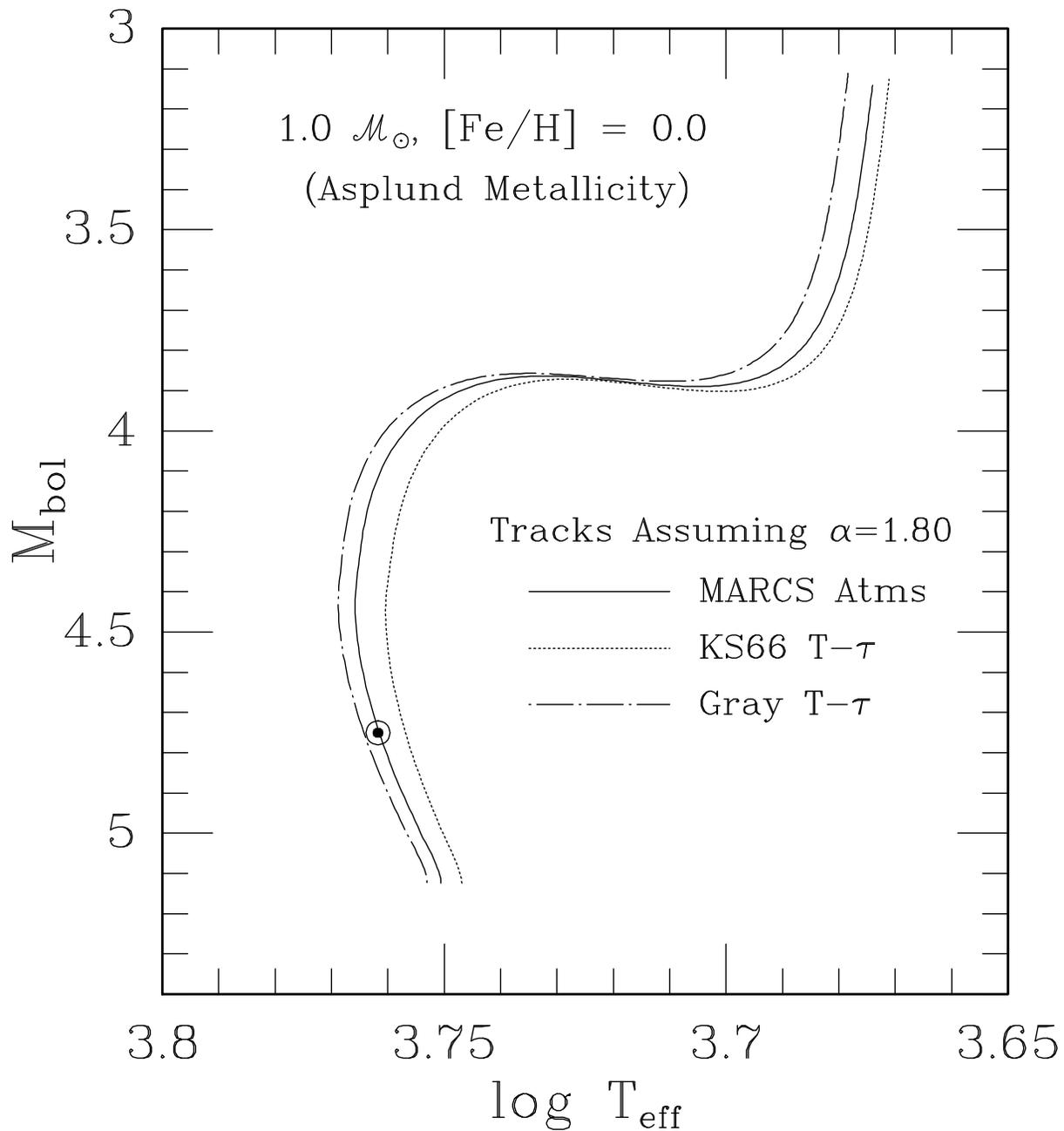}
\caption{Similar to the previous figure, except that the tracks for the 
different treatments of the atmospheric layers all assume $\amlt = 1.80$.  Since
the tracks using MARCS atmospheres differ only slightly when attached at the
photosphere or at $\tau = 100$, only the former case is plotted here.}
\label{fig:vanfig4}
\end{figure}

\clearpage
\begin{figure}
\plotone{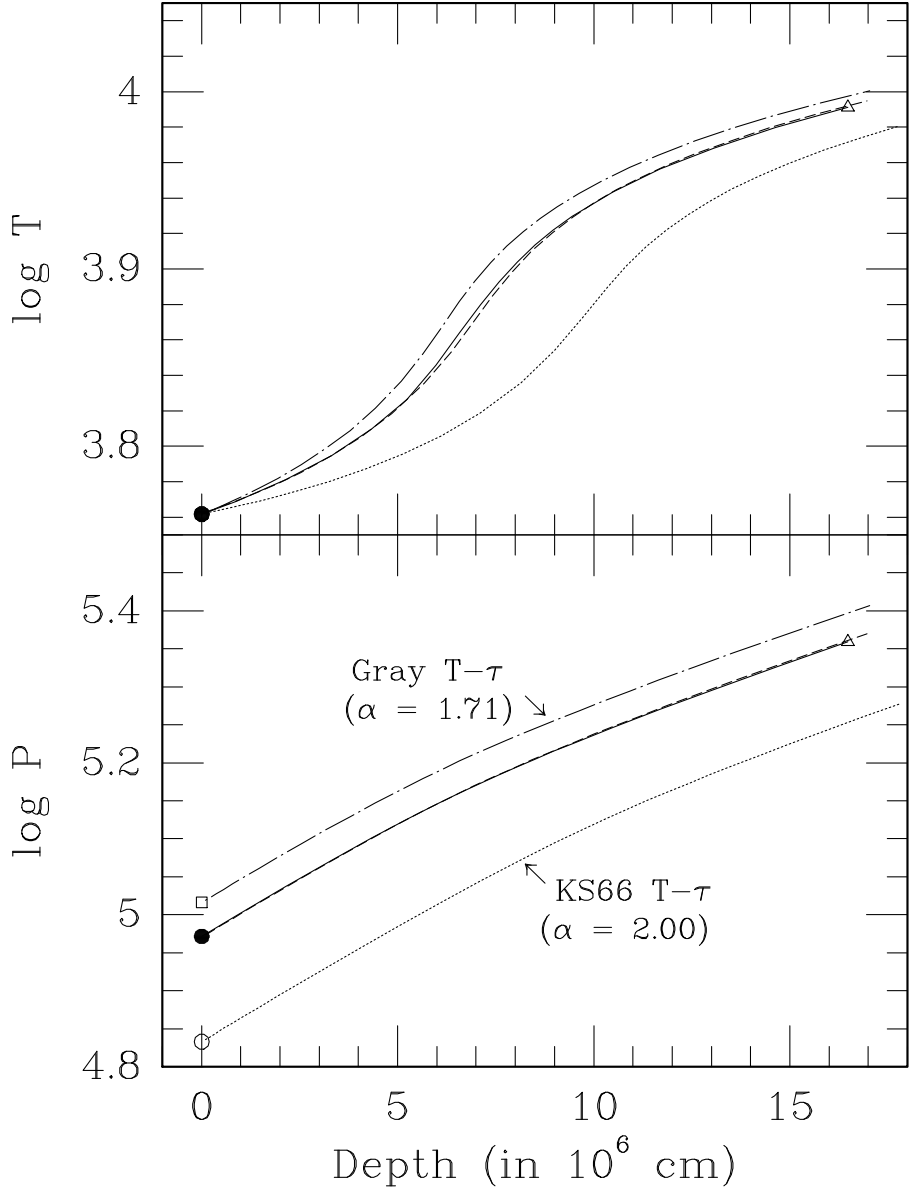}
\caption{Plot of the variations of $T$ and $P$ with geometrical depth in the
sub-photospheric layers of Standard Solar Models.  (In this and similar plots,
the depth is defined to be zero at the photosphere.)  The solid curve represents
a MARCS model atmosphere for the Sun.  The dashed, dot-dashed, and dotted
curves indicate the results of Runge-Kutta integrations (see the text) in which
the initial value of $T$ is the observed $\teff$ of the Sun and the initial
values of $P$ were obtained from the MARCS model or by integrating the
hydrostatic equation for the gray or KS66 $\fttau$ relations, respectively.
The solid and dashed curves both assume $\amlt = 1.80$: the adopted
mixing-length parameters for the other two cases are as indicated.  The symbols
attached to the various curves provide identifications for the integrations to
deeper layers that are plotted in Figure~\ref{fig:vanfig7}: one of these
integrations begins at the location of the open triangle, which indicates the
layer in the MARCS solar atmosphere where $\tau = 100$.}
\label{fig:vanfig5}
\end{figure}

\clearpage
\begin{figure}
\plotone{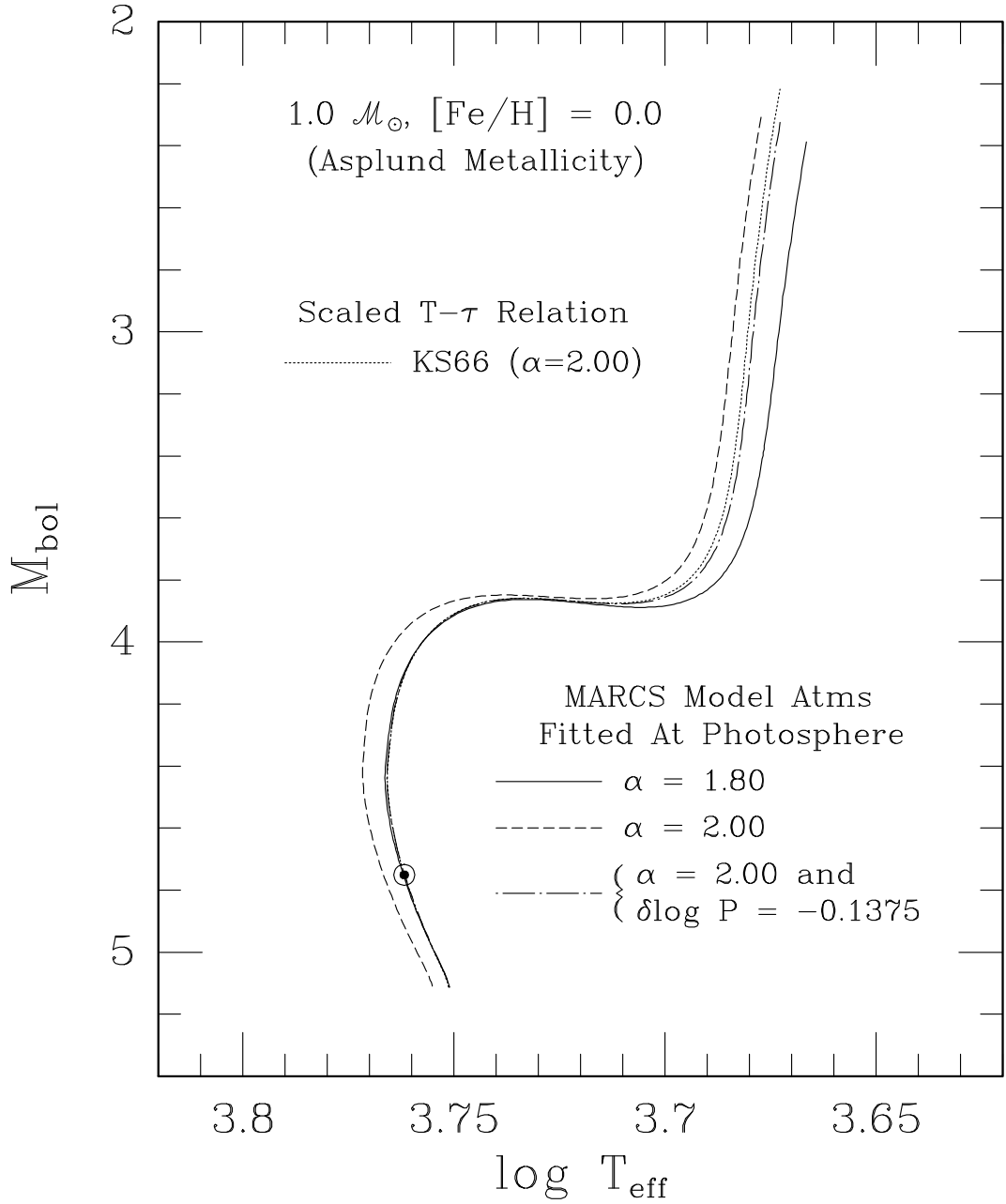}
\caption{Similar to Fig.~\ref{fig:vanfig3}, except that the effects on the
tracks of increasing $\amlt$ from 1.80 to 2.00, and of adjusting the 
photospheric pressures predicted the MARCS atmospheres for $\amlt = 2.0$
by $\delta\log P = -0.1375$ are shown.  Both the dotted and solid curves are
identical to their counterparts in Fig.~\ref{fig:vanfig3}.}
\label{fig:vanfig6}
\end{figure}
 
\clearpage
\begin{figure}
\plotone{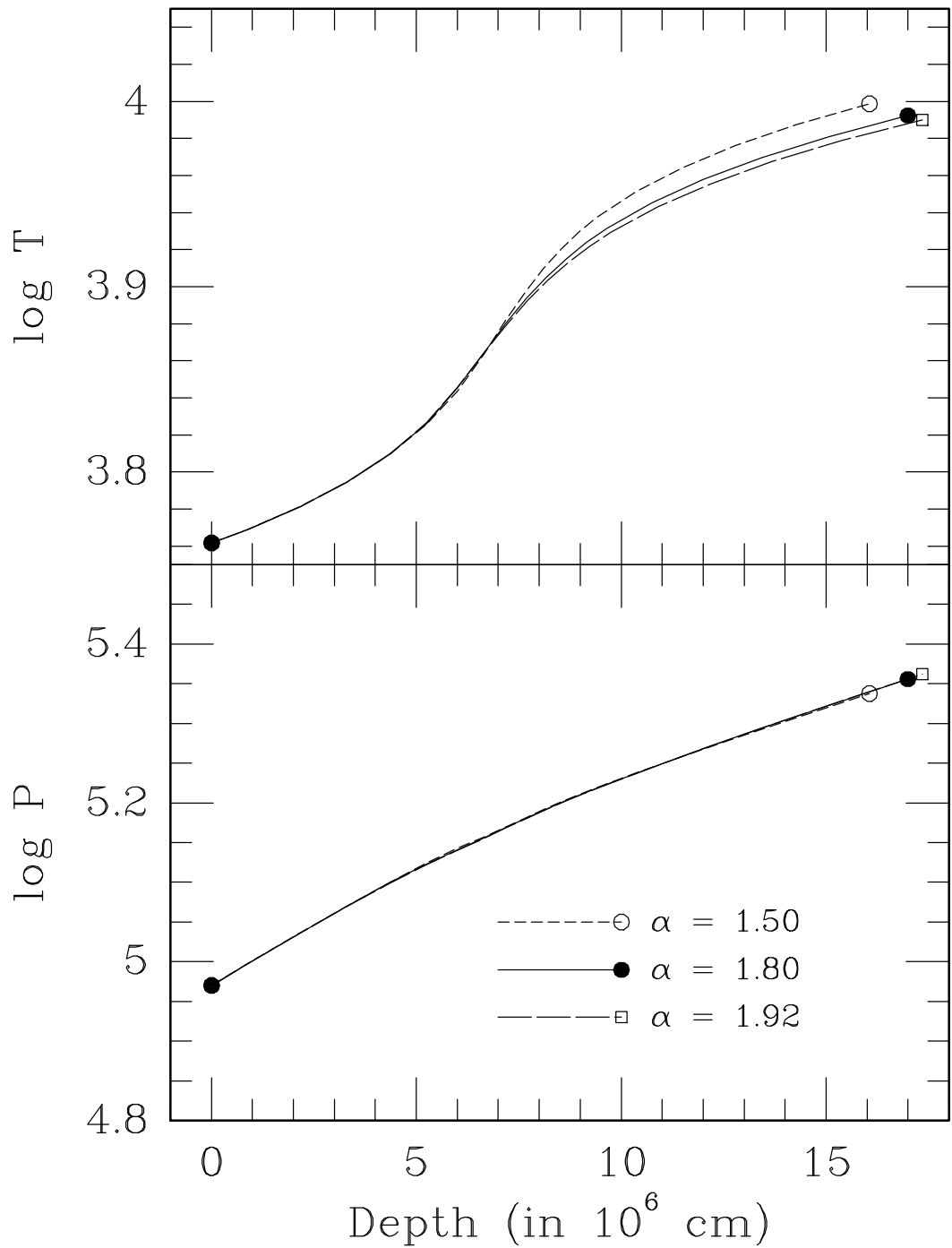}
\caption{The variations of $log T$ and $\log P$ with depth between the 
photosphere and that layer where $\tau = 100$ from MARCS model atmospheres for
$\log g = 4.44$, $\teff = 5777$ K, Asplund abundances, and the three indicated
values of $\amlt$.  The solid curve plotted here is the same as the solid
curve plotted in the previous figure.}
\label{fig:vanfig7}
\end{figure}

\clearpage
\begin{figure}
\plotone{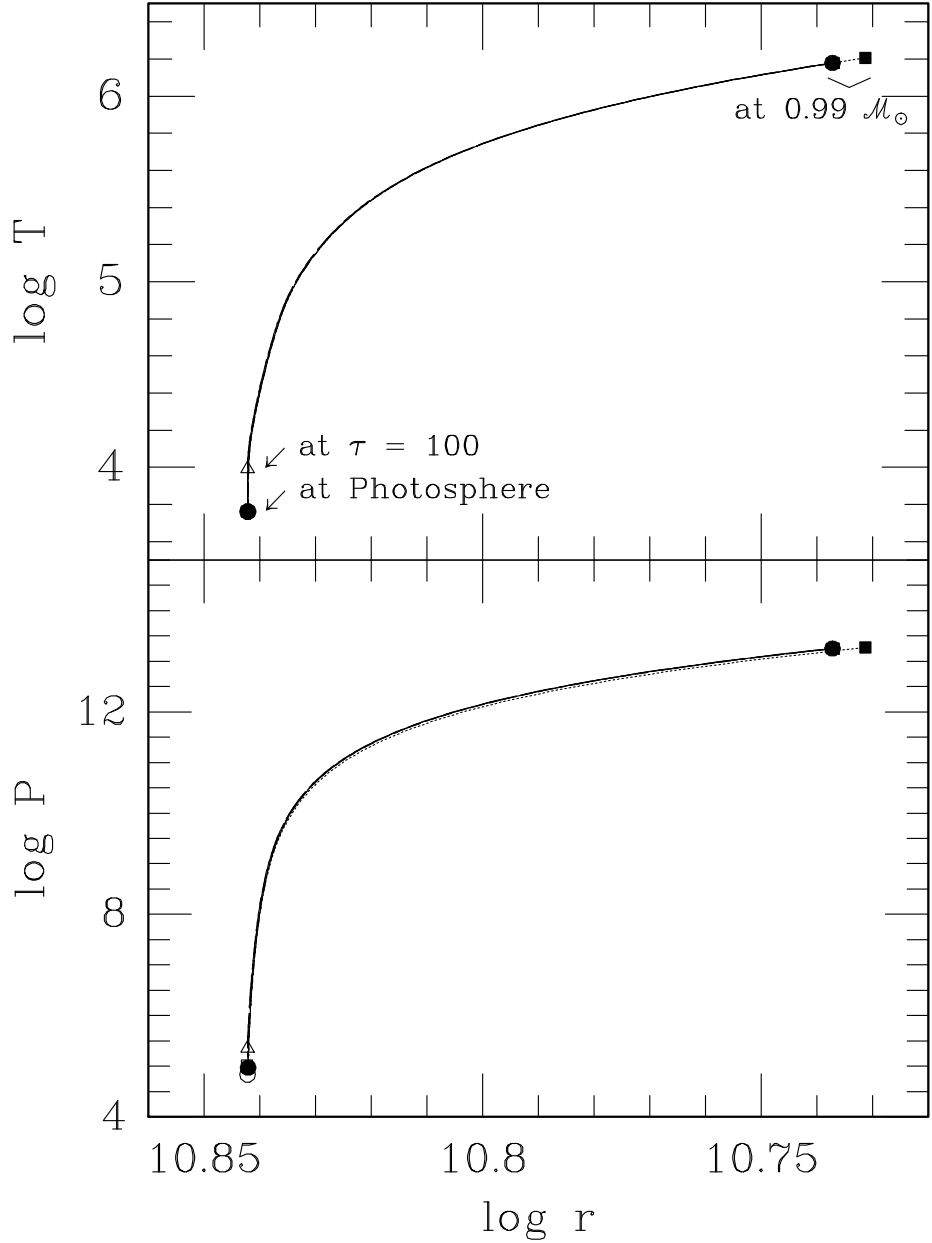}
\caption{Similar to Fig.~\ref{fig:vanfig5}, except that the Runge-Kutta
integrations are extended to $0.99 \msol$ (where the outer boundary conditions
used to solve the stellar structure equations are defined; see the text).  The
symbols, which identify the same cases presented in Fig.~\ref{fig:vanfig5}, are
plotted at both the beginning and end-points of the integrations: most are not
visible because they superimpose one another.  The dotted curve and filled
square were obtained by repeating the integration represented by the long-dashed
curve and filled circle plotted in Fig.~\ref{fig:vanfig5}, but assuming $\amlt =
1.71$ instead of 1.80.}
\label{fig:vanfig8}
\end{figure} 

\clearpage
\begin{figure}
\plotone{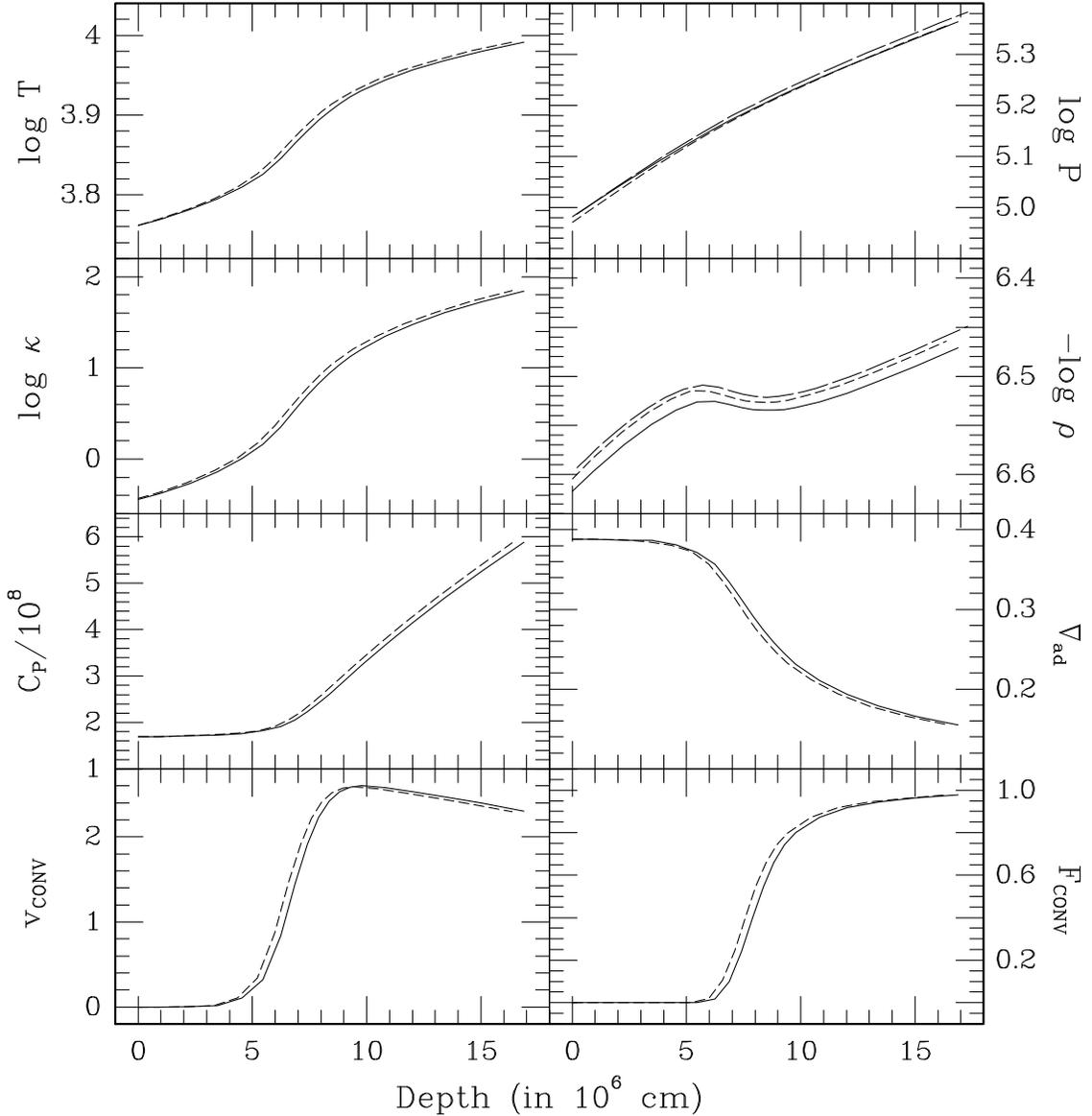}
\caption{Similar to Fig.~\ref{fig:vanfig1}; in this case, the predictions from
MARCS model atmospheres, with and without taking macroturbulence into account
(solid and dashed curves, respectively) are compared.  Long-dashed curves in
the pressure-depth and density-depth panels indicate the results of Runge-Kutta
integrations inward from the photophere with the initial value of the pressure
taken from the turbulent model atmosphere.  Because these integrations, as
performed by the Victoria code, do not allow for macroturbulence, the
long-dashed curves tend to be close to the dashed curves.  In fact, they are so
similar in the case of all of the other quantities considered in this figure
that the long-dashed curves have not been plotted in the other panels (for the
sake of clarity).  All of the calculations assume $\amlt=1.77$, and solar model
parameters.}
\label{fig:vanfig9}
\end{figure}

\clearpage
\begin{figure}
\plotone{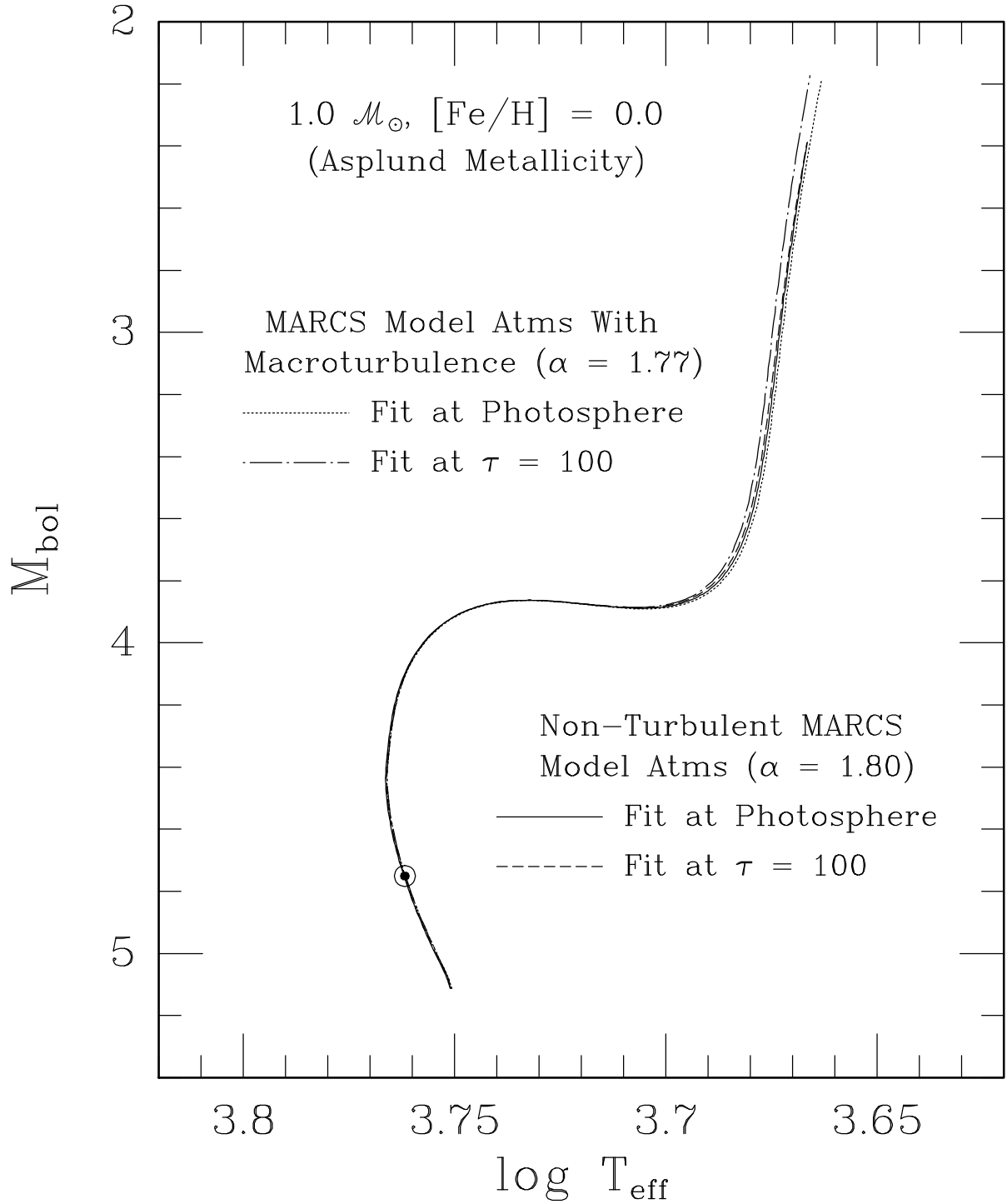}
\caption{Similar to Fig.~\ref{fig:vanfig3}; in this case, evolutionary tracks
are compared that use turbulent or non-turbulent MARCS model atmospheres as
boundary conditions.}
\label{fig:vanfig10}
\end{figure}

\clearpage
\begin{figure}
\plotone{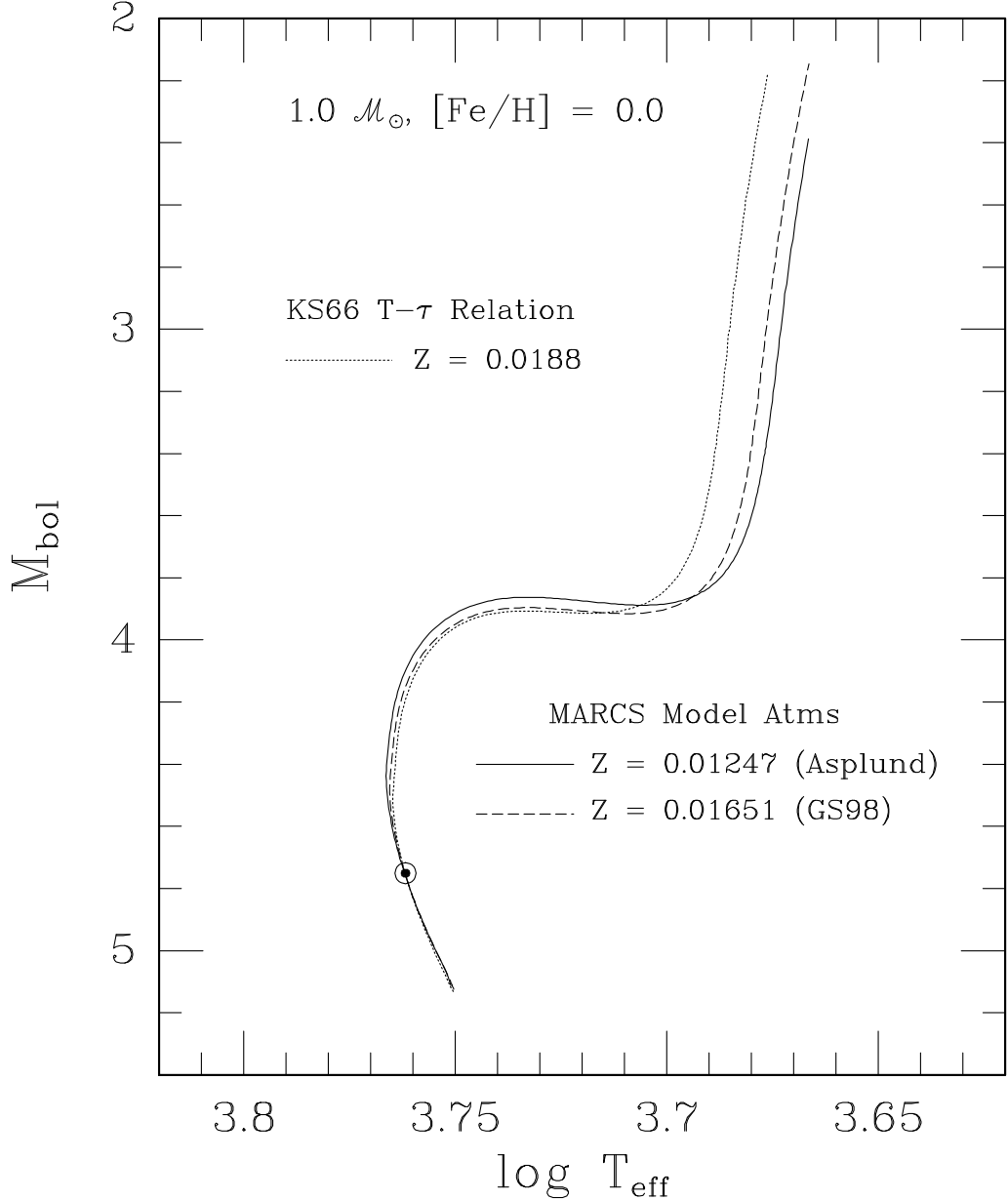}
\caption{Similar to Fig.~\ref{fig:vanfig3}; in this case, the dependence of
solar evolutionary tracks on the assumed value of $Z_\odot$ is shown.  The solid
and dashed curves were computing using boundary conditions derived from
non-turbulent MARCS model atmospheres for $Z_\odot = 0.01247$ and 0.01651, on
the assumption of the Asplund and GS98 heavy-element mixtures, respectively,
and the values of $Y$ and $\amlt$ given in Table~\ref{tab:tab2}.  They were
fitted to the interior models at the photosphere.  The dotted curve was taken
from the grid of models computed by VandenBerg et al.~(2006): it assumes
$Z_\odot = 0.0188$ (with the heavy-element mixture given by \citealt{gn93}),
$Y_\odot = 0.2715$, and $\amlt = 1.91$.  Boundary pressures for these models
were determined using KS66 atmospheric structures.}
\label{fig:vanfig11}
\end{figure}

%
%

\clearpage
\begin{figure}
\plotone{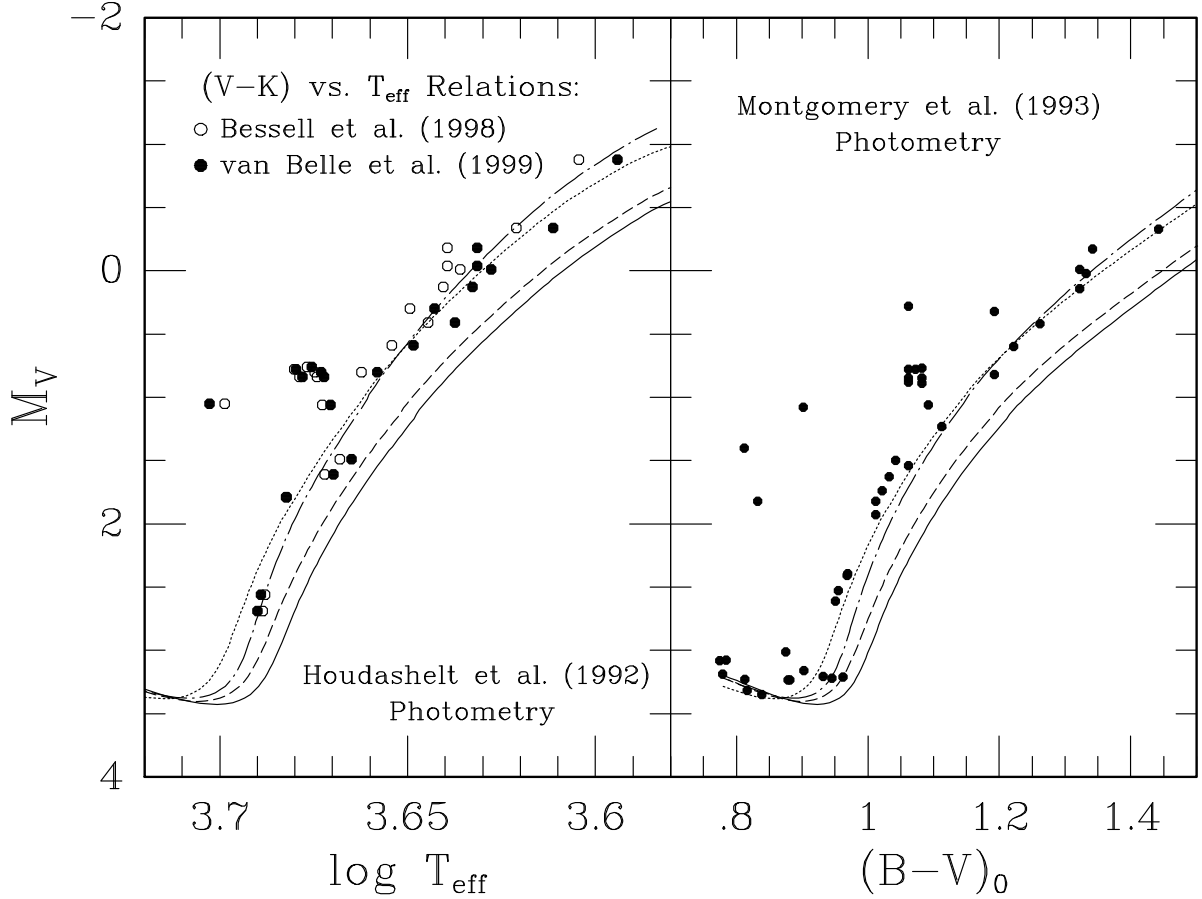}
\caption{{\it Left-hand panel}: Plot of the M$\,$67 giants for which
temperatures were derived by applying the empirical $(V-K)$--$\teff$ relations
by Bessell et al.~(1998; open circles) and van Belle et al.~(1999; filled
circles) to the $VK$ photometry by Houdashelt et al.~(1992).  The assumed
$E(B-V)$ and $(m-M)_V$ values are 0.038 mag and 9.70 mag, respectively.  The
dotted curve gives the giant-branch segment of the 4.0 Gyr, $Z=0.0173$ isochrone
that was used by VandenBerg \& Clem (2003) to fit the M$\,$67 CMD.  The solid
and dashed curves represent, in turn, the extensions to high luminosities of the
4.2 Gyr, $Z=0.01247$ and 3.9 Gyr, $Z=0.01651$ isochrones that were fitted to
the CMD of M$\,$67 by VandenBerg et al.~(2007).  The dot-dashed curve
represents an otherwise identical calculation to that indicated by the solid
curve, except that MARCS SDC model atmospheres were used as boundary conditions
(see the text).  {\it Right-hand panel}: As for the left-hand panel, except
that the models are compared with the $BV$ photometry of cluster giants 
reported by Montgomery et al.~(1993).}
\label{fig:vanfig12}
\end{figure}

\clearpage
\begin{figure}
\plotone{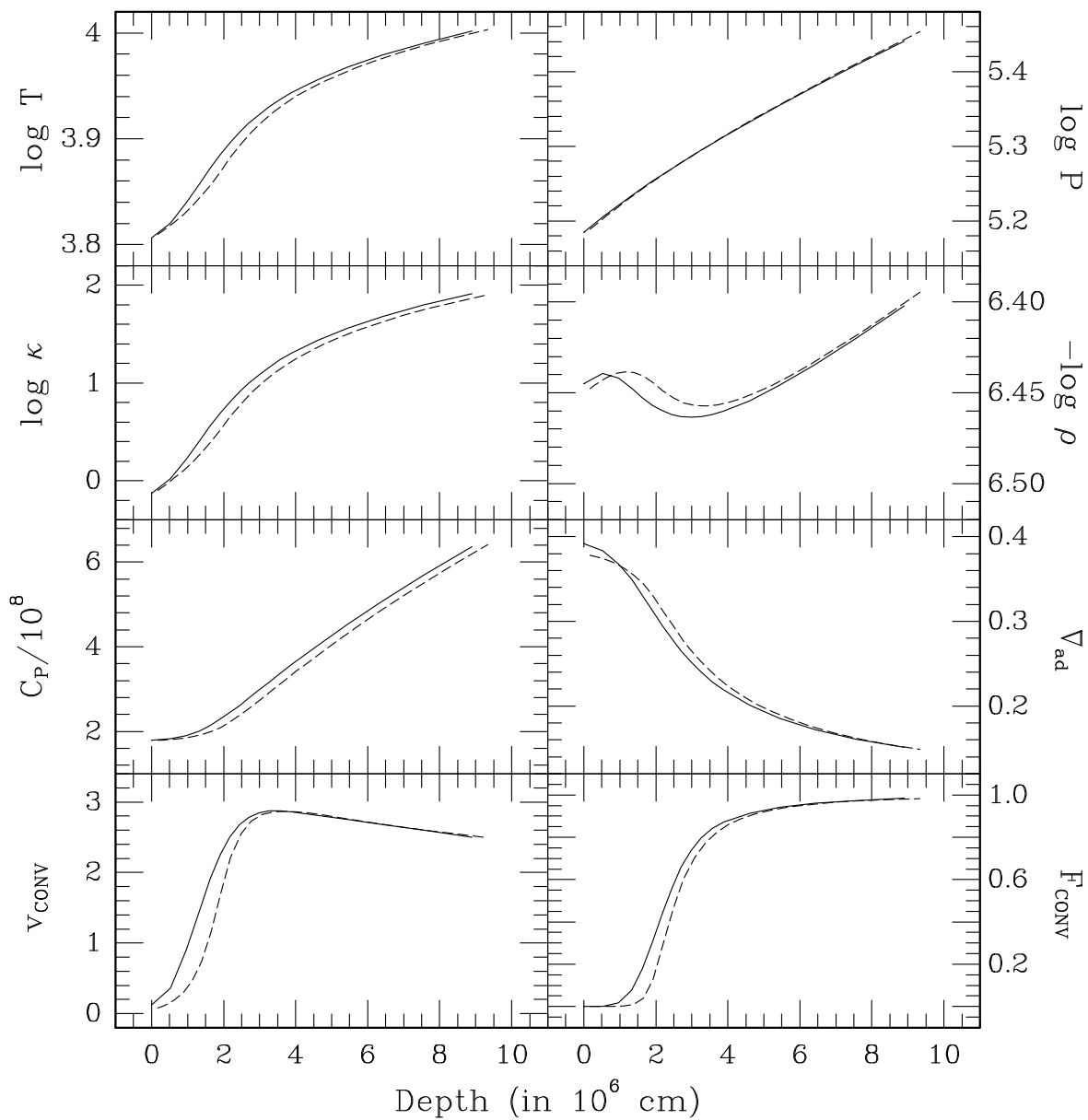}
\caption{Similar to Fig.~\ref{fig:vanfig1}, except that the comparison of 
the various quantities is made for a model having $\teff = 6398$ K, $\log g =
4.59$, and [Fe/H] $= -2.0$ (assuming the heavy-element mixture in
Table~\ref{tab:tab3}).}
\label{fig:vanfig13}
\end{figure}

\clearpage
\begin{figure}
\plotone{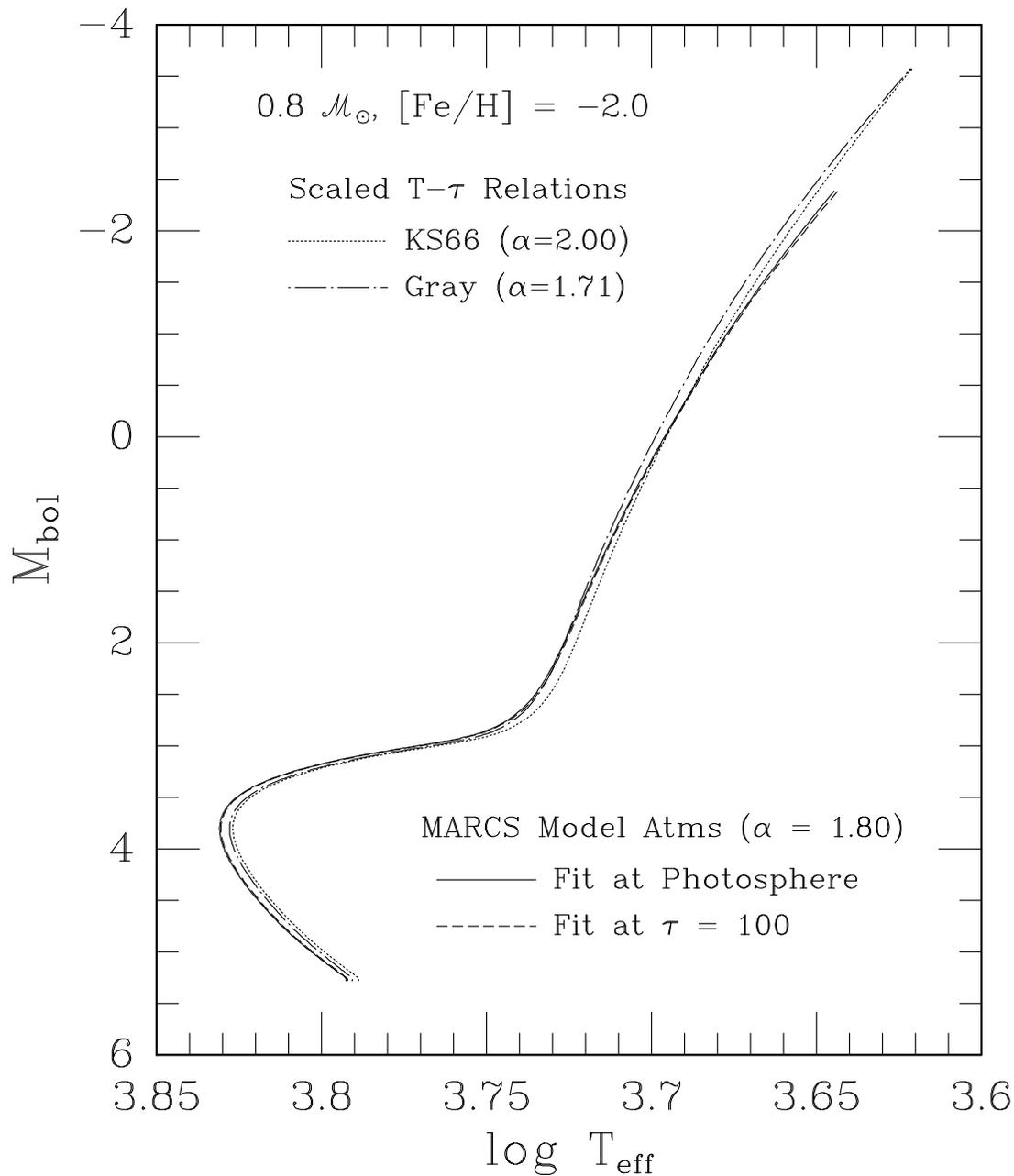}
\caption{Similar to Fig.~\ref{fig:vanfig3}, except that tracks for $0.8 \msol$
stars having $Y=0.24823$ and $Z=0.00028$ (assuming the heavy-element mixture
in Table~\ref{tab:tab3}) are plotted.  The indicated values of $\amlt$ are the
same as those required by the corresponding Standard Solar Models.  The solid
and dashed curves do not extend to the tip of the giant branch because MARCS
model atmospheres were not computed for gravities less than $\log g = 0.0$.}
\label{fig:vanfig14}
\end{figure}

\clearpage
\begin{figure}
\plotone{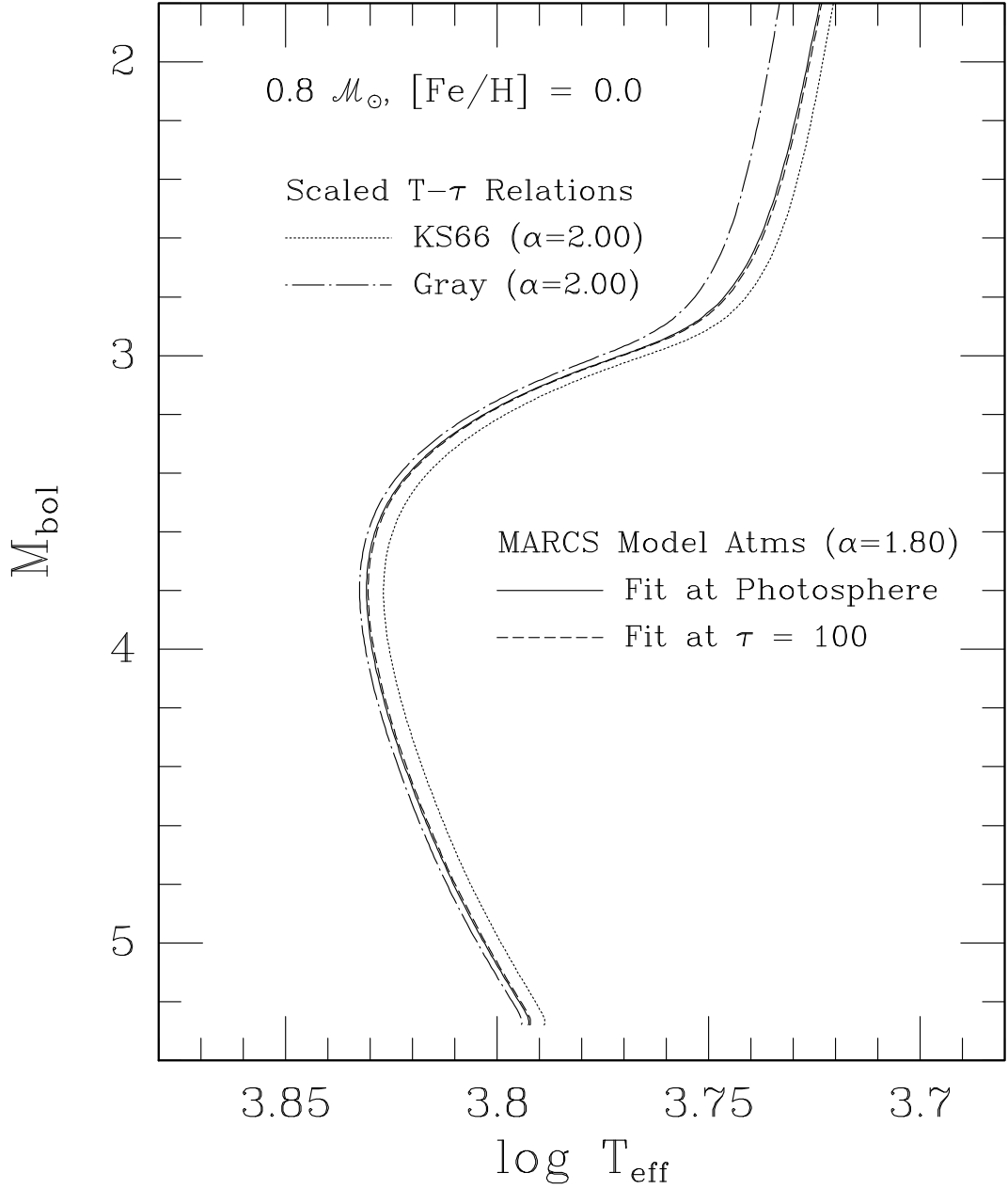}
\caption{An expanded version of the previous plot (to reveal the differences in
the vicinity of the turnoff more clearly) in which the track using gray
atmospheres has been recomputed on the assumption of $\amlt = 2.0$ (instead of
1.71).}
\label{fig:vanfig15}
\end{figure}

\begin{figure}
\plotone{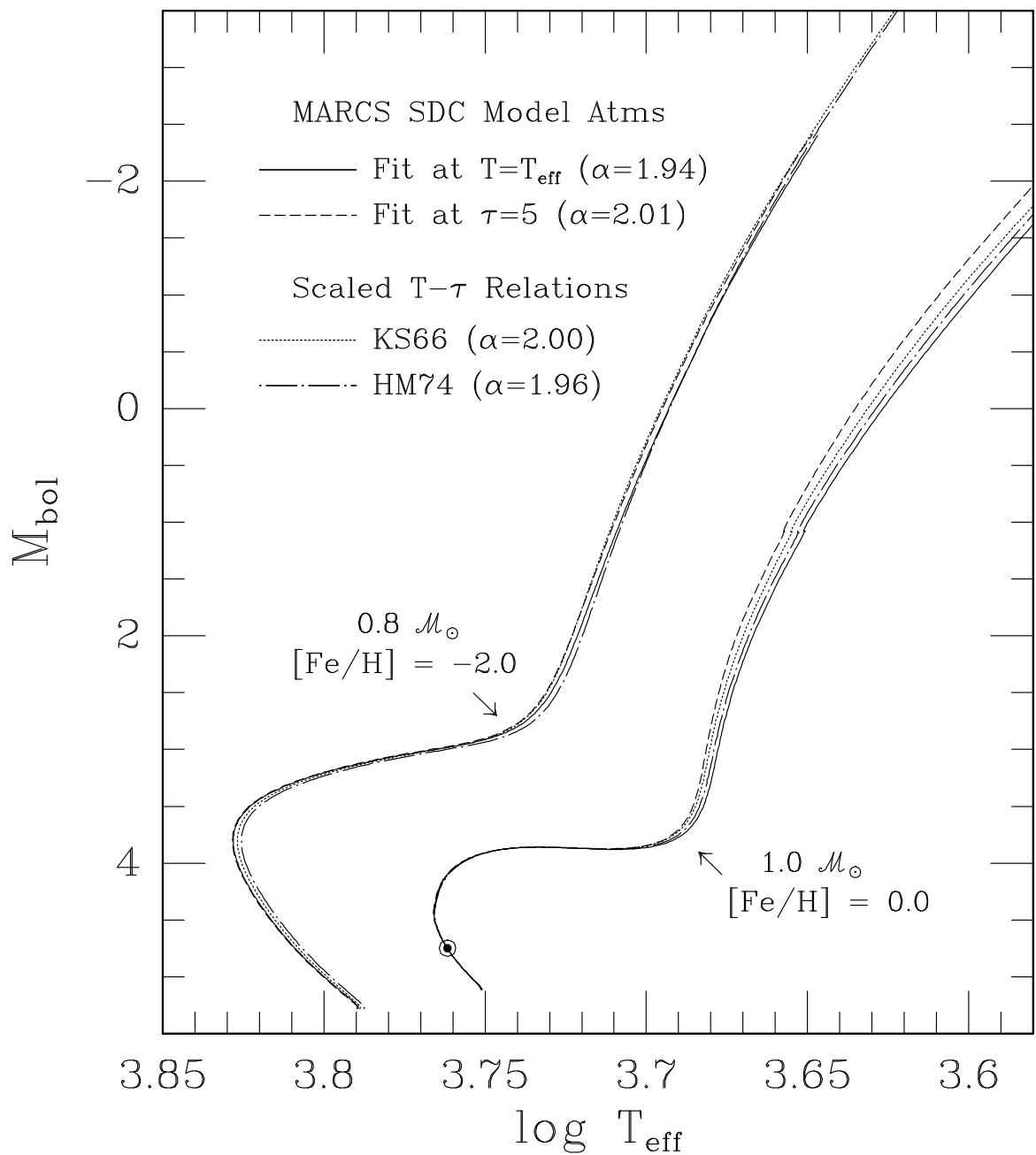}
\caption{Comparisons of the tracks for $1.0 {{\cal M}_\odot}$, solar-abundance
stars and those for $0.8 {{\cal M}_\odot}$ stars having [Fe/H] $=-2.0$ when four
different treatments of the atmospheric layers are assumed, as indicated.  The
mixing-length parameter has been set, in each case, to the value required by
a Standard Solar Model.}
\label{fig:vanfig16}
\end{figure}

\begin{figure}
\plotone{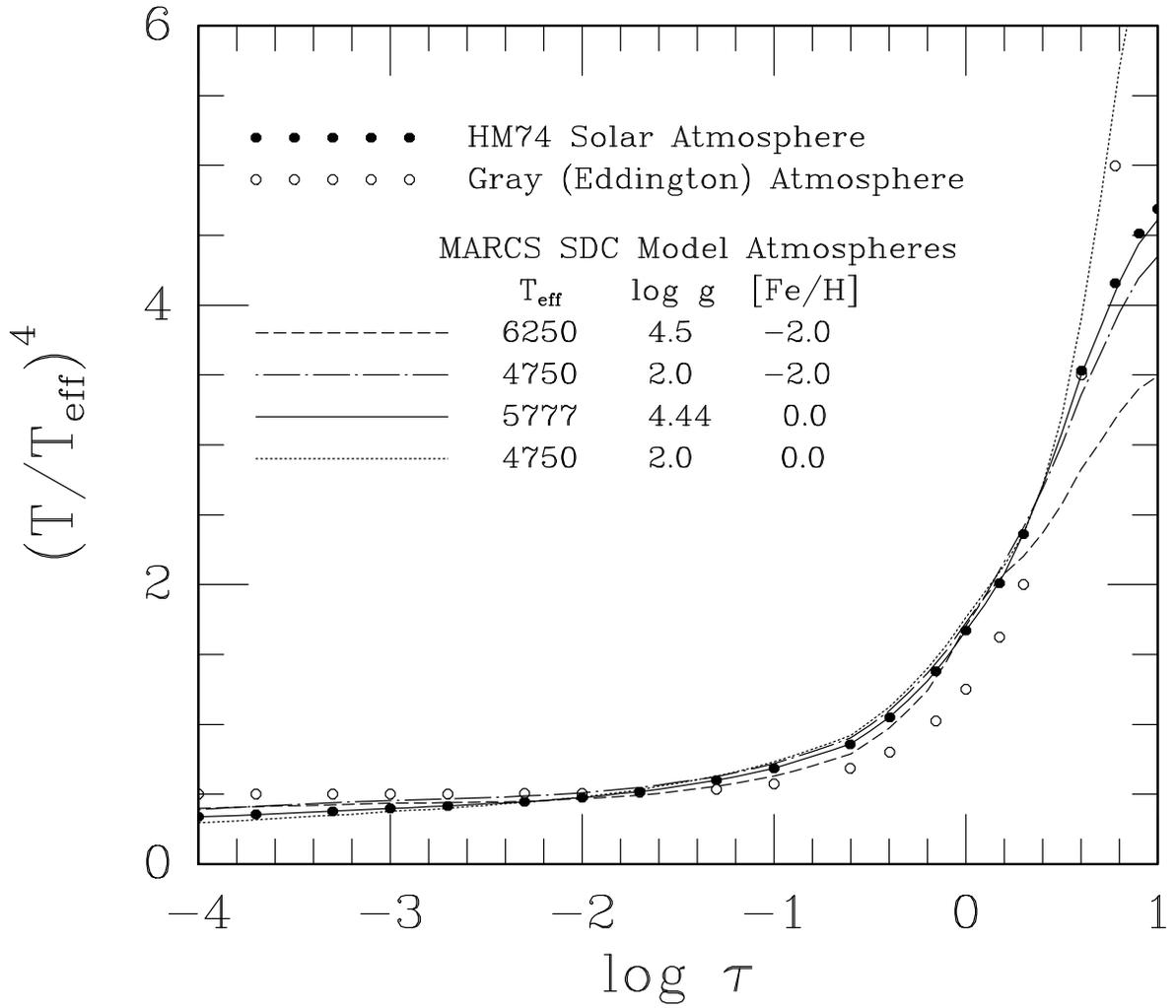}
\caption{Comparison of the scaled $T$--$\tau$ structures predicted by MARCS
SDC model atmospheres for the indicated parameters with those given by the
HM74 and classical gray atmospheres.}
\label{fig:vanfig17}
\end{figure} 

\clearpage
\begin{figure}
\plotone{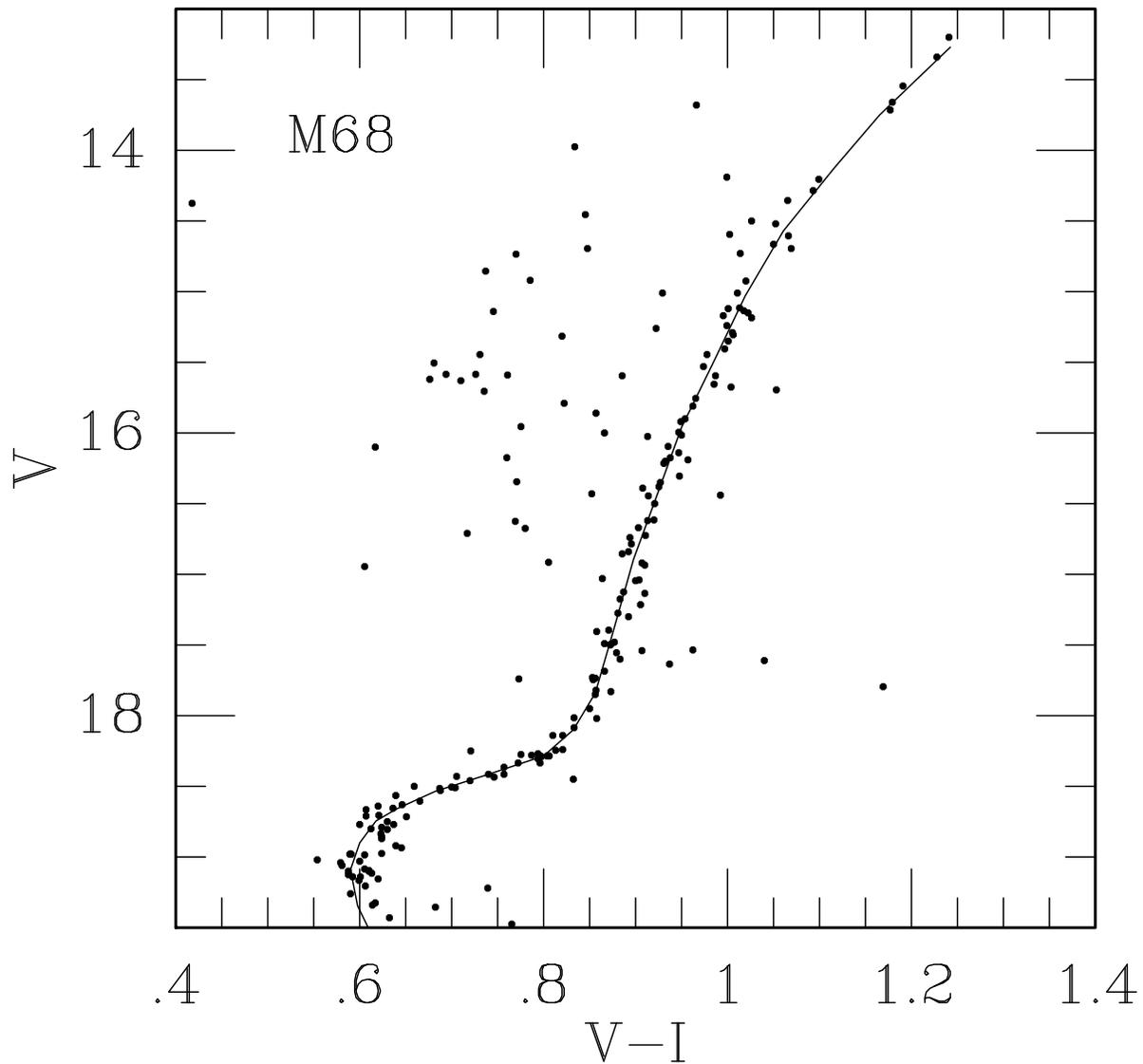}
\caption{Plot of the standard field photometry of the globular cluster M$\,$68
provided by Stetson (2000; small filled circles) with the fiducial sequence
(solid curve) derived in this investigation to represent the observed CMD for
the evolved stars.  This merges smoothly into the mean locus for the fainter
stars derived by VandenBerg (2000) from the $VI$ photometry obtained by
Walker (1994).  The combined fiducial so obtained is used in the following
figures.}
\label{fig:vanfig18}
\end{figure}

\clearpage
\begin{figure}
\plotone{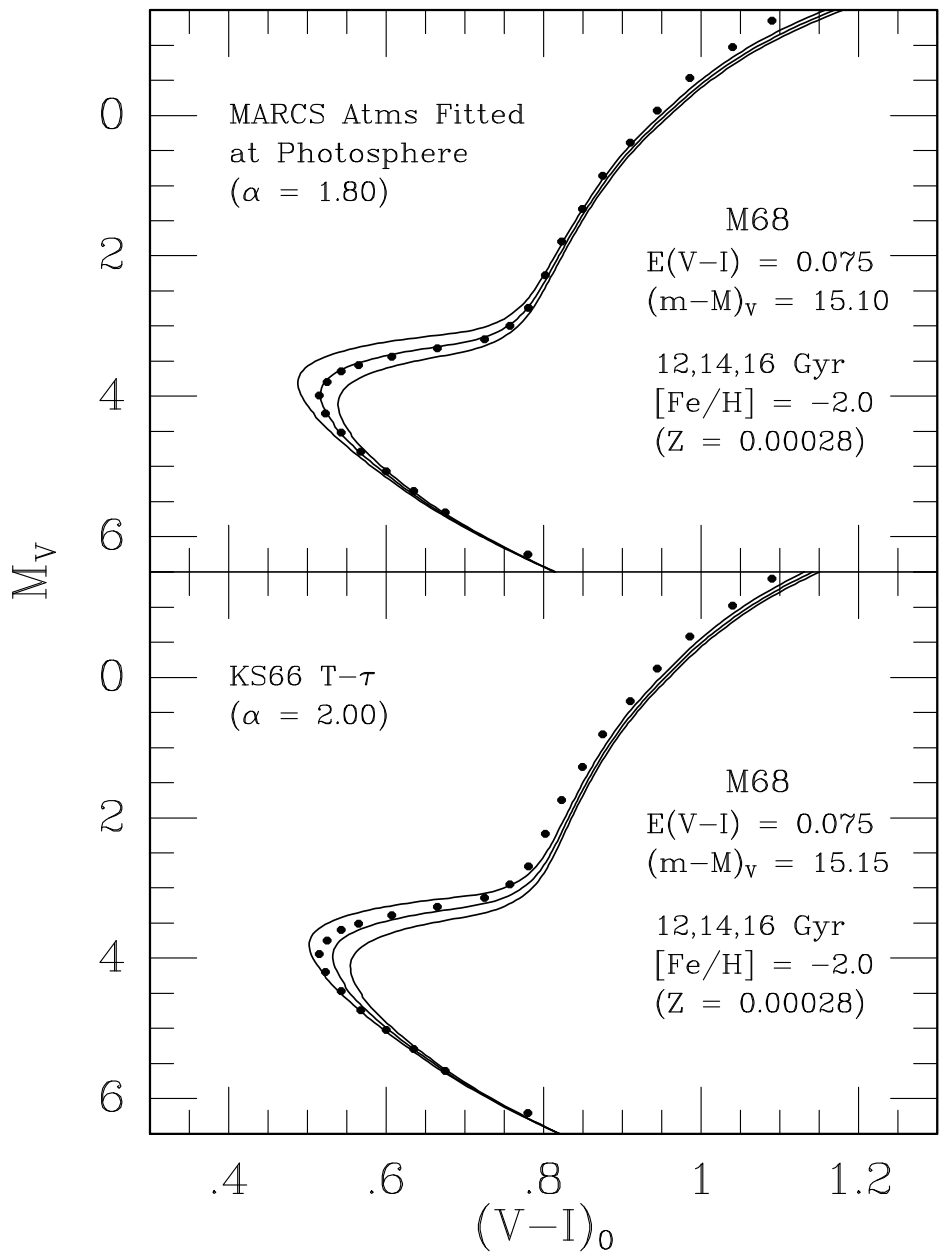}
\caption{Comparisons of isochrones (solid curves) for the indicated ages and
chemical abundances with the adopted fiducial sequence of M$\,$68 (filled
circles).  The models used in the upper and lower panels are based on different
treatments of the atmosphere and values values of $\amlt$, as noted.  The
adopted reddening is from the Schlegel et al.~(1998) dust maps, while the
indicated distance moduli were derived from main-sequence fits to the 
isochrones.}
\label{fig:vanfig19}
\end{figure}

\clearpage
\begin{figure}
\plotone{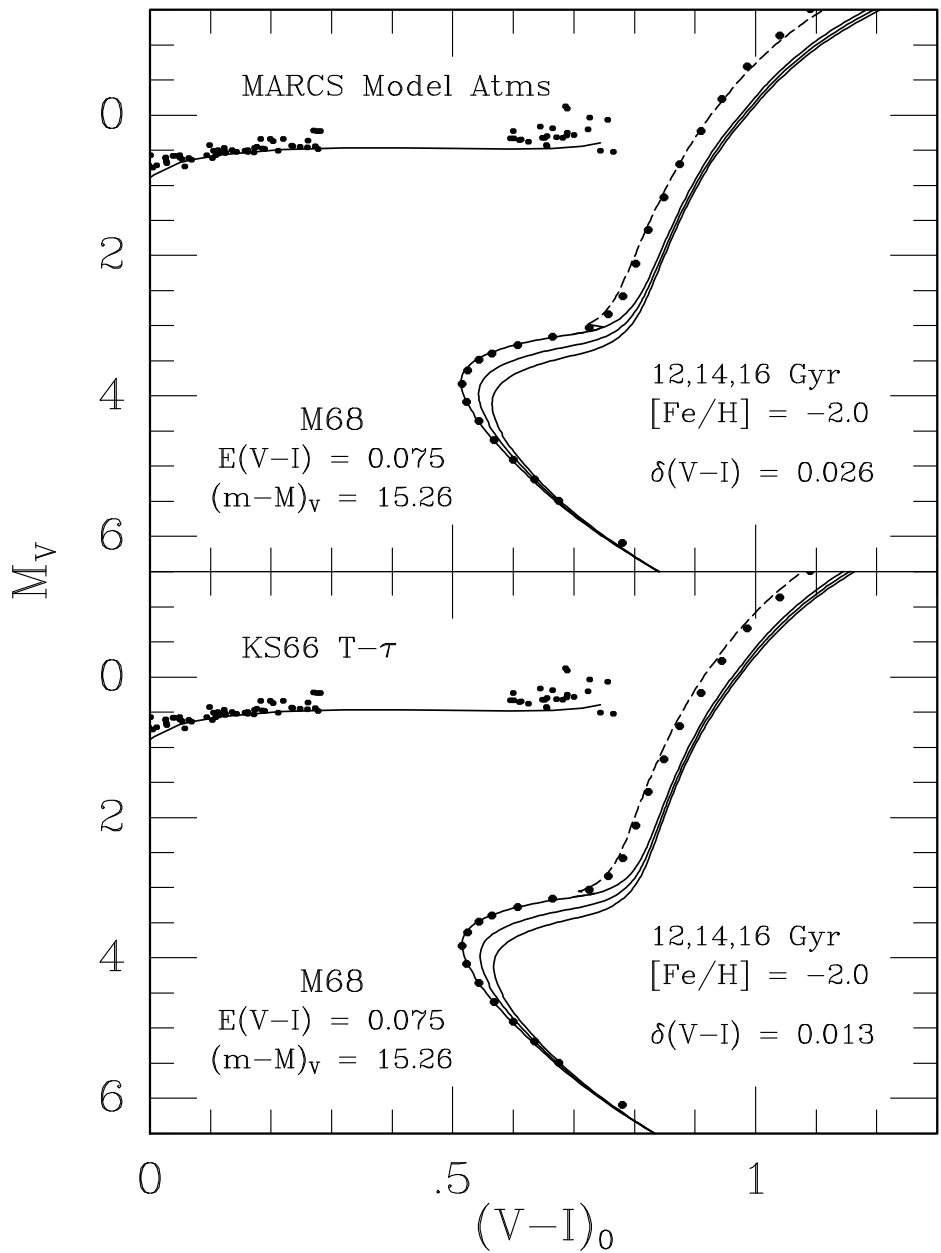}
\caption{Similar to the previous figure, except that the assumed distance is
based on a fit of a fully consistent ZAHB to the lower bound of the distribution
of the cluster horizontal-branch stars.  The source of the HB photometry is
Walker (1994).  Note that the isochrones had to be shifted to the red, by the
indicated amounts, to achieve a superposition of their lower main-sequence
segments with the cluster fiducial at $M_V \gta 5$.}
\label{fig:vanfig20}
\end{figure}

\clearpage
\begin{deluxetable}{ccc}
\tabletypesize{\footnotesize}
\tablecaption{Solar Elemental Abundances \label{tab:tab1}}
\tablewidth{0pt}
\tablehead{ & \multispan2\hfil $\log N$\hfil \\ & \multispan2\hrulefill \\ 
Element & Grevesse \& Sauval (1998) & Asplund\tablenotemark{a} } 
\startdata
C\phantom{e} & \phantom{1}8.52 & \phantom{1}8.41 \\
N\phantom{e} & \phantom{1}7.92 & \phantom{1}7.80 \\
O\phantom{e} & \phantom{1}8.83 & \phantom{1}8.66 \\
Ne           & \phantom{1}8.08 & \phantom{1}7.84 \\
Na           & \phantom{1}6.33 & \phantom{1}6.33 \\
Mg           & \phantom{1}7.58 & \phantom{1}7.58 \\
Al           & \phantom{1}6.47 & \phantom{1}6.47 \\
Si           & \phantom{1}7.55 & \phantom{1}7.51 \\
P\phantom{e} & \phantom{1}5.45 & \phantom{1}5.45 \\
S\phantom{e} & \phantom{1}7.33 & \phantom{1}7.33 \\
Cl           & \phantom{1}5.50 & \phantom{1}5.50 \\
Ar           & \phantom{1}6.40 & \phantom{1}6.18 \\
K\phantom{e} & \phantom{1}5.12 & \phantom{1}5.12 \\
Ca           & \phantom{1}6.36 & \phantom{1}6.36 \\
Ti           & \phantom{1}5.02 & \phantom{1}5.02 \\
Cr           & \phantom{1}5.67 & \phantom{1}5.67 \\
Mn           & \phantom{1}5.39 & \phantom{1}5.39 \\
Fe           & \phantom{1}7.50 & \phantom{1}7.45 \\
Ni           & \phantom{1}6.25 & \phantom{1}6.25 \\
\multispan3\hrulefill \\
$Z$        & \phantom{1}0.0165 & \phantom{1}0.0125 \\
\enddata
\tablenotetext{a}{Abundances for C, N, O, Ne, Si, Ar, and Fe were provided
by M.~Asplund (2004, priv.~comm.): Grevesse \& Sauval (1998) abundances are
assumed for all other elements heavier than helium.
($\log N_{\rm H} \equiv 12.00$.)}
\end{deluxetable}

\clearpage
\begin{deluxetable}{cccccc}
\tabletypesize{\footnotesize}
\tablecaption{Parameters of Standard Solar Models \label{tab:tab2}}
\tablewidth{0pt}
\tablehead{ Metals Mixture & $\amlt$\tablenotemark{a} &
 $\log N_{\rm He}$ & $X$ & $Y$ & $Z$ }
\startdata
Grevesse \& Sauval (1998) & 1.84 & 10.9738 & 0.71585 & 0.26764 & 0.01651 \\
Asplund (see Table 1)     & 1.80 & 10.9487 & 0.72995 & 0.25758 & 0.01247 \\
\enddata
\tablenotetext{a}{Obtained when MARCS non-turbulent model atmospheres are
attached to the interior structures at $T = \teff$.}
\end{deluxetable}

\clearpage
\begin{deluxetable}{ccccccc}
\tabletypesize{\footnotesize}
\tablecaption{Adopted Abundances at [Fe/H] $= -2$ \label{tab:tab3}}
\tablewidth{0pt}
\tablehead{ Element & & $\log N$ & & [m/H] & & [m/Fe] }
\startdata
C\phantom{e}& ~~~~ & 6.41 & ~~~~ & $-2.0$ & ~~~~ & \phantom{$+$}$0.0$ \\
N\phantom{e}& ~~~~ & 5.80 & ~~~~ & $-2.0$ & ~~~~ & \phantom{$+$}$0.0$ \\
O\phantom{e}& ~~~~ & 7.16 & ~~~~ & $-1.5$ & ~~~~ &             $+0.5$ \\
Ne          & ~~~~ & 6.14 & ~~~~ & $-1.7$ & ~~~~ &             $+0.3$ \\
Na          & ~~~~ & 4.03 & ~~~~ & $-2.3$ & ~~~~ &             $-0.3$ \\
Mg          & ~~~~ & 5.88 & ~~~~ & $-1.7$ & ~~~~ &             $+0.3$ \\
Al          & ~~~~ & 4.47 & ~~~~ & $-2.0$ & ~~~~ & \phantom{$+$}$0.0$ \\
Si          & ~~~~ & 5.91 & ~~~~ & $-1.6$ & ~~~~ &             $+0.4$ \\
P\phantom{e}& ~~~~ & 3.45 & ~~~~ & $-2.0$ & ~~~~ & \phantom{$+$}$0.0$ \\
S\phantom{e}& ~~~~ & 5.63 & ~~~~ & $-1.7$ & ~~~~ &             $+0.3$ \\
Cl          & ~~~~ & 3.50 & ~~~~ & $-2.0$ & ~~~~ & \phantom{$+$}$0.0$ \\
Ar          & ~~~~ & 4.48 & ~~~~ & $-1.7$ & ~~~~ &             $+0.3$ \\
Ca          & ~~~~ & 4.66 & ~~~~ & $-1.7$ & ~~~~ &             $+0.3$ \\
Ti          & ~~~~ & 3.22 & ~~~~ & $-1.8$ & ~~~~ &             $+0.2$ \\
Cr          & ~~~~ & 3.27 & ~~~~ & $-2.4$ & ~~~~ &             $-0.4$ \\
Mn          & ~~~~ & 2.99 & ~~~~ & $-2.4$ & ~~~~ &             $-0.4$ \\
Fe          & ~~~~ & 4.45 & ~~~~ & $-2.0$ & ~~~~ & \phantom{$+$}$0.0$ \\
Ni          & ~~~~ & 4.25 & ~~~~ & $-2.0$ & ~~~~ & \phantom{$+$}$0.0$ \\
\enddata
\end{deluxetable}


\newpage
 
\begin{thebibliography}{}

\bibitem[Asplund(2004)]{asp04}
Asplund, M.~2004, in Proceedings of JD4 at the XXIII IAU General Assembly:
Astrophysical Impact of Abundances in Globular Clusters, eds.~F.~D'Antona \&
G.~Da Costa, Mem.~Soc.~Astron.~Ital., 75, 300.

\bibitem[Asplund, Grevesse, \& Sauval(2006)]{ags06}
Asplund, M., Grevesse, N., \& Sauval, A.~J.~2006, Comm.~in Asteroseismology,
147, 76.

\bibitem[Asplund et al.(2005)]{ags05}
Asplund, M., Grevesse, N., Sauval, S.~J., Allende Prieto, C., \& Blomme, 
R.~2005, A\&A, 431, 693.

\bibitem[Asplund, Nordlund, \& Trampedach(1999)]{ant99}
Asplund, M., Nordlund, A., \& Trampedach, R.~1999, In Theory and Tests of
Convection in Stellar Structure, eds.~A.~Gim\'enez, E.~F.~Guinan, \&
B.~Montesinos, ASP Conf.~Ser., 173, 221.

\bibitem[Baraffe et al.(1997)]{bca97}
Baraffe, I., Chabrier, G., Allard, F., \& Hauschildt, P.~H.~1997, A\&A, 327, 
1054.

\bibitem[Bell \& Gustafsson(1989)]{bg89}
Bell, R.~A., \& Gustafsson, B.~1989, MNRAS, 236, 653.

\bibitem[Bessell \& Brett(1988)]{bb88}
Bessell, M.~S., \& Brett, J.~M.~1988, PASP, 100, 1134.

\bibitem[Bessell, Castelli, \& Plez(1998)]{bcp98}
Bessell, M.~S., Castelli, F., \& Plez, B.~1998, A\&A, 333, 231.

\bibitem[Blackwell, Lynas-Gray, \& Smith(1995)]{bls95}
Blackwell, D.~E., Lynas-Gray, A.~E., \& Smith, G.~1995,A\&A, 296, 217.

\bibitem[B\"ohm-Vitense(1958)]{bv58}
B\"ohm-Vitense, E.~1958, Zs.~Ap., 46, 108.

\bibitem[Brocato, Cassisi, \& Castellani(1998)]{bcc98}
Brocato, E., Cassisi, S., \& Castellani, V.~1998, MNRAS, 295, 711.

\bibitem[Cacciari, Corwin, \& Carney(2005)]{ccc05}
Cacciari, C., Corwin, T.~M., \& Carney, B.~W.~2005, AJ, 129, 267.

\bibitem[Carretta \& Gratton(1997)]{cg97}
Carretta, E., \& Gratton, R.~G.~1997, A\&AS, 121, 95.

\bibitem[Cassisi et al.(2004)]{csc04}
Cassisi, S., Salaris, M., Castelli, F., \& Pietrinferni, A.~2004, ApJ, 616, 498.

\bibitem[Castelli(1999)]{cas99}
Castelli, F.~1999, A\&A, 346, 564.

\bibitem[Cayrel et al.(2004)]{cds04}
Cayrel, R., et al.~2004, A\&A, 416, 1117.

\bibitem[Coc et al.(2004)]{cvd04}
Coc, A., Vangioni-Flam, E., Descouvemont, P., Adahchour, A., \& Angulo, C.~2004,
ApJ, 600, 544.

\bibitem[Collet et al.(2007)]{cat07}
Collet R., Asplund M., \& Trampedach R.~2007, astro-ph/0703652.

\bibitem[De Santis \& Cassisi(1999)]{dc99}
De Santis, R., \& Cassisi, S.~1999, MNRAS, 308, 97.

\bibitem[Ferguson et al.(2005)]{faa05}
Ferguson, J.~W., Alexander, D.~R., Allard, F., Barman, T., Bodnarik, J.~G.,
Hauschildt, P.~H., Heffner-Wong, A., \& Tamani, A.~2005, ApJ, 623, 585.
 
\bibitem[Ferraro et al.(2006)]{fvs06}
Ferraro, F.~R., Valenti, E., Straniero, O., \& Origlia, L.~2006, ApJ, 642, 225.

\bibitem[Girardi et al.(2000)]{gbb00}
Girardi, L, Bressan, A., Bertelli, G., \& Chiosi, C.~2000, A\&AS, 141, 371.

\bibitem[Grevesse \& Noels(1993)]{gn93}
Grevesse, N., \& Noels, A.~1993, Phys.~Scr., T47, 133.

\bibitem[Grevesse \& Sauval(1998)]{gs98}
Grevesse, N., \& Sauval, A.~J.~1998, Space Sci.~Rev., 85, 161.\ \ \ \ (GS98)

\bibitem[Gustafsson et al.(1975)]{gbe75}
Gustafsson, B., Bell, R.~A., Eriksson, K., \& Nordlund, A.~1975, A\&A, 42. 407.

\bibitem[Gustafsson et al.(2003)]{gee03}
Gustafsson, B., Edvardsson, B.,  Eriksson, K., Mizuno-Wiedner, M., 
Jorgensen, U.~G., \& Plez, B.~2003, in Stellar Atmosphere Modelling,
eds.~I.~Hubeny, D.~Mihalas, \& K.~Werner, ASP Conf.~Ser., 288, 331.

\bibitem[Henyey, Forbes, \& Gould(1964)]{hfg64}
Henyey, L.~G., Forbes, J.~E., \& Gould, N.~L.~1964, ApJ, 139, 306.

\bibitem[Heiter \& Eriksson(2006)]{he06}
Heiter, U., \& Eriksson, K.~2006, A\&A, 452, 1039.

\bibitem[Henyey, Vardya, \& Bodenheimer(1965)]{hvb65}
Henyey, L., Vardya, M.~S., \& Bodenheimer, P.~1965, ApJ, 142, 841.

\bibitem[Hobbs \& Thorburn(1991)]{ht91}
Hobbs, L.~M., \& Thorburn, J.~A.~1991, AJ, 102, 1070.

\bibitem[Holweger \& M\"uller(1974; hereafter HM74)]{hm74}
Holweger, H., \& M\"uller, E.~A.~1974, Solar Phys., 39, 19.\ \ \ \ (HM74)

\bibitem[Houdashelt, Bell, \& Sweigart(2000)]{hbs00}
Houdashelt, M.~L., Bell, R.~A., \& Sweigart, A.~V.~2000, AJ, 119, 1448.

\bibitem[Houdashelt, Frogel, \& Cohen(1992)]{hfc92}
Houdashelt, M.~L, Frogel, J.~A., \& Cohen, J.~G.~1992, AJ, 103, 163.

\bibitem[Hubbard \& Lampe(1969)]{hl69}
Hubbard, W.~B., \& Lampe, M.~1969, ApJS, 18, 297.

\bibitem[Iglesias \& Rogers(1996)]{ir96}
Iglesias, C.~A., \& Rogers, F.~J.~1996, ApJ, 464, 943.

\bibitem[Irwin(1985)]{ir85}
Irwin, A.~W.~1985, A\&A, 146, 282.

\bibitem[Kippenhahn, Weigert, \& Hofmeister(1967)]{kwh67}
Kippenhahn, R., Weigert, A., \& Hofmeister, E.~1967, in Methods of
Computational Physics, vol.~7, eds.~B.~Alder, S.~Fernbach, \& M.~Rotenberg (New
York: Academic), p. 129.

\bibitem[Korn et al.(2006)]{kgr06}
Korn, A.~J., Grundahl, F., Richard, O., Barklem, P.~S., Mashonkina, L., 
Collet, R., Piskunov, N., \& Gustafsson, B.~2006, Nature, 442, 657.

\bibitem[Kraft \& Ivans(2003)]{ki03}
Kraft, R.~P., \& Ivans, I.~I.~2003, PASP, 115, 143.
 
\bibitem[Krishna Swamy(1966)]{ks66}
Krishna Swamy, K.~S.~1966, ApJ, 145, 174.\ \ \ \ (KS66)

\bibitem[Lejeune, Cuisinier, \& Buser(1998)]{lcb98}
Lejeune, T., Cuisinier, F., \& Buser, R.~1998, A\&AS, 130, 65.

\bibitem[Ludwig, Freytag, \& Steffen(1999)]{lfs99}
Ludwig, H.G., Freytag, B., \& Steffen, M.~1999, A\&A, 346, 111.
 
\bibitem[Mihalas(1970)]{mi70}
Mihalas, D.~1970, {\it Stellar Atmospheres} (San Francisco: W.~H.~Freeman).

\bibitem[Montalb\'an et al.(2001)]{mkd01}
Montalb\'an, J., Kupka, F., D'Antona, F., \& Schmidt, W.~2001, A\&A, 370, 982.

\bibitem[Montgomery, Marschall, \& Janes(1993)]{mmj93}
Montgomery, K.~A., Marschall, L.~A., \& Janes, K.~A.~1993, AJ, 106, 181.

\bibitem[Morel et al.(1994)]{mvp94}
Morel, P., van't Meer, C., Provost, J., Berthomieu, G., Castelli, F., Cayrel,
R., Goupil, M.~J., \& Lebreton, Y.~1994, A\&A, 286, 91.

\bibitem[Pedersen, VandenBerg, \& Irwin(1990)]{pvi90}
Pedersen, B.~B., VandenBerg, D.~A., \& Irwin, A.~W.~1990, ApJ, 352, 279.

\bibitem[Pietrinferni et al.(2004)]{pcs04}
Pietrinferni, A., Cassisi, S., Salaris, M., \& Castelli, F.~2004, ApJ, 612, 168.

\bibitem[Potekhin et al.(1999)]{pbh99}
Potekhin, A.~Y., Baiko, D.~A., Haensel, P., \& Yakovlev, D.~G.~1999, A\&A, 346,
345.

\bibitem[Randich et al.(2006)]{rsp06}
Randich, S., Sestito, P., Primas, F., Pallavicini, R., \& Pasquini, L.~2006,
A\&A, 450, 557. 

\bibitem[Richard et al.(2002)]{rmr02}
Richard, O., Michaud, G., Richer, J., Turcotte, S., Turck-Chi\'eze, \&
VandenBerg, D.~A.~2002, ApJ, 568, 979.

\bibitem[Rood \& Crocker(1985)]{rc85}
Rood, R.~T., \& Crocker, D.~A.~1985, in Production and Distribution of C, N, O
Elements, eds.~I.~J.~Danziger, F.~Matteucci, \& K.~Kj\"ar (Garching: ESO),
p.~61.

\bibitem[Salaris, Cassisi, \& Weiss(2002)]{scw02}
Salaris, M., Cassisi, S., \& Weiss, A.~2002, PASP, 114, 375.

\bibitem[Schlegel, Finkbeiner, \& Davis(1998)]{sfd98}
Schlegel, D.~J., Finkbeiner, D.~P., \& Davis, M.~1998, ApJ, 500, 525.
 
\bibitem[Sekiguchi \& Fukugita(2000)]{sf00}
Sekiguchi, M., \& Fukugita, M.~2000, AJ, 120, 1072.

\bibitem[Stetson(2000)]{st00}
Stetson, P.~B.~2000, PASP, 112, 925.

\bibitem[Tautvai{\u s}ien{\.e} et al.(2000)]{tet00}
Tautvai{\u s}ien{\.e}, G., Edvardsson, B., Tuominen, I., \& Ilyin, I.~2000,
A\&A, 360, 499.

\bibitem[van Belle et al.(1999)]{vb99}
van Belle, G.~T., et al.~1999, AJ, 117, 521.
 
\bibitem[VandenBerg(1983)]{van83}
VandenBerg, D.~A.~1983, ApJS, 51, 29.

\bibitem[VandenBerg(1991)]{van91}
VandenBerg, D.~A.~1991, in ASP.~Conf.~Ser.~13, The Formation and
Evolution of Star Clusters, ed.~K.~Janes (San Francisco: ASP), 183.

\bibitem[VandenBerg(1992)]{van92}
VandenBerg, D.~A.~1992, ApJ, 392, 685.

\bibitem[VandenBerg(2000)]{van00}
VandenBerg, D.~A.~2000, ApJS, 129, 315.

\bibitem[VandenBerg, Bergbusch, \& Dowler(2006)]{vbd06}
VandenBerg, D.~A., Bergbusch, P.~A., \& Dowler, P.~D.~2006, ApJS, 162, 375.
 
\bibitem[VandenBerg \& Clem(2003)]{vc03}
VandenBerg, D.~A., \& Clem, J.~L.~2003, AJ, 126, 778.

\bibitem[VandenBerg et al.(2007)]{vge07}
VandenBerg, D.~A., Gustafsson, B., Edvardsson, B.,  Eriksson, K., \& Ferguson,
J.~2007, ApJL, submitted.

\bibitem[VandenBerg \& Poll(1989)]{vp89}
VandenBerg, D.~A., \& Poll, H.~E.~1989, AJ, 98, 1451.

\bibitem[VandenBerg et al.(2002)]{vrm02}
VandenBerg, D.~A., Richard, O., Michaud, G., \& Richer, J.~2002, ApJ, 571, 487.

\bibitem[VandenBerg et al.(2000)]{vsr00}
VandenBerg, D.~A., Swenson, F.~J., Rogers, F.~J., Iglesias, C.~A., \&
Alexander, D.~R.~2000, ApJ, 532, 430.

\bibitem[Walker(1994)]{wal94}
Walker, A.~R.~1994, AJ, 108, 555.

\bibitem[Yi et al.(2001)]{ydk01}
Yi, S., Demarque, P.,  Kim, Y.-C., Lee, Y.-W., Ree, C.~H., Lejeune, T., \&
Barnes, S.~2001, ApJS, 136, 417.
 
\end{thebibliography}
\end{document}